\documentclass[prx,reprint,aps,amsmath,amssymb,amsfonts,superscriptaddress,notitlepage,floatfix]{revtex4-1}

\usepackage{pifont}

\usepackage[colorlinks=true,citecolor=blue,urlcolor=blue,linkcolor=blue,hyperfigures=true]{hyperref}
\usepackage{graphicx}
\usepackage{amsmath,amsfonts}
\usepackage[table,x11names]{xcolor}
\usepackage{color}
\usepackage{changes}

\newcommand{\up}{\uparrow}
\newcommand{\dn}{\downarrow}
\newcommand{\dv}{\ensuremath{\mathbf{d}}}

\newcommand{\rv}{\ensuremath{\mathbf{r}}}
\newcommand{\Rv}{\ensuremath{\mathbf{R}}}
\newcommand{\kv}{\ensuremath{\mathbf{k}}}
\newcommand{\qv}{\ensuremath{\mathbf{q}}}
\newcommand{\Qv}{\ensuremath{\mathbf{Q}}}
\newcommand{\Kv}{\ensuremath{\mathbf{K}}}

\newcommand{\av}[1]{\ensuremath{\left\langle #1 \right\rangle}}

\newcommand{\abs}[1]{\ensuremath{\left| #1 \right|}}

\renewcommand{\Re}{\operatorname{Re}}

\definecolor{ErikBlauw}{cmyk}{1.,0.3,0.,0.} 
\definecolor{ErikRood}{cmyk}{.2,0.9,0.8,0.} 
\definecolor{ErikGroen}{cmyk}{1,0.20,1,0} 

\definecolor{ErikOranje}{HTML}{F68542}
\definecolor{ErikGrijs}{HTML}{868586}
\definecolor{ErikGrijs2}{HTML}{AAAAAA}
\definecolor{ErikWit}{HTML}{FFFFFF}
\definecolor{ErikZwart}{HTML}{000000}

\definecolor{erik_red}{RGB}{215,25,28}
\definecolor{erik_redb}{RGB}{255,113,110}
\definecolor{erik_blueb}{RGB}{5,113,176}
\definecolor{erik_yellow}{RGB}{255,255,191}
\definecolor{erik_white}{RGB}{255, 255, 255}

\definecolor{ourGreen}{HTML}{009E73}
\definecolor{ourPurple}{HTML}{9400D3}

\usepackage{tikz}
\usetikzlibrary{calc}
\usetikzlibrary{arrows}
\usetikzlibrary{patterns}
\usetikzlibrary{decorations.pathmorphing}
\usetikzlibrary{decorations.pathreplacing}
\usetikzlibrary{decorations.markings}
\usetikzlibrary{shapes}
\usetikzlibrary{shapes.geometric}
\usetikzlibrary{positioning}

\begin{document}

\title{Random Phase Approximation for gapped systems: role of vertex corrections and applicability of the constrained random phase
approximation}

 \author{Erik G. C. P. van Loon}
 \altaffiliation{Currently at Department of Physics,  Lund University, Professorsgatan 1, 223 63, Lund, Sweden}
 \email{erik.van\_loon@teorfys.lu.se}
 \affiliation{Institut f{\"u}r Theoretische Physik, Universit{\"a}t Bremen, Otto-Hahn-Allee 1, 28359 Bremen, Germany}
 \affiliation{Bremen Center for Computational Materials Science, Universit{\"a}t Bremen, Am Fallturm 1a, 28359 Bremen, Germany}

 \author{Malte R{\"o}sner}
 \affiliation{Institute for Molecules and Materials, Radboud University, Heyendaalseweg 135, NL-6525AJ Nijmegen, The Netherlands}
 
 \author{Mikhail I. Katsnelson}
 \affiliation{Institute for Molecules and Materials, Radboud University, Heyendaalseweg 135, NL-6525AJ Nijmegen, The Netherlands}
 
 \author{Tim O. Wehling}
 \affiliation{Institut f{\"u}r Theoretische Physik, Universit{\"a}t Bremen, Otto-Hahn-Allee 1, 28359 Bremen, Germany}
 \affiliation{Bremen Center for Computational Materials Science, Universit{\"a}t Bremen, Am Fallturm 1a, 28359 Bremen, Germany}

\begin{abstract}
The many-body theory of interacting electrons poses an intrinsically difficult problem that requires simplifying assumptions. 
For the determination of electronic screening properties of the Coulomb interaction, the Random Phase Approximation (RPA) provides such a simplification. 
Here, we explicitly show that this approximation is justified for band structures with sizeable band gaps. This is when the electronic states responsible for the screening are energetically far away from the Fermi level, which is equivalent to a short electronic propagation length of these states. 
The RPA contains exactly those diagrams in which the classical Coulomb interaction covers all distances, whereas neglected vertex corrections involve quantum tunneling through the barrier formed by the band gap.
Our analysis of electron-electron interactions provides a real-space analogy to Migdal's theorem on the smallness of vertex corrections in electron-phonon problems.
An important application is the increasing use of constrained Random Phase Approximation (cRPA) calculations of effective interactions.
We find that their usage of Kohn-Sham energies already accounts for the leading local (excitonic) vertex correction in insulators.
\end{abstract}

\maketitle

The Random Phase Approximation (RPA) plays an important role in condensed matter theory. Introduced by Bohm and Pines in the 1950s~\cite{Bohm51,Bohm52,Bohm53}, it provides a self-consistent, microscopic view on the Coulomb interaction between electrons. Nowadays, the approximation is used in \emph{ab initio} methods to calculate energetics~\cite{Schimka10}, dielectric properties~\cite{Shishkin07}, plasmon spectra~\cite{Hwang07}, the polarizability of molecules~\cite{Jorgensen20} and solids and the effective interaction strengths~\cite{Aryasetiawan04} in low-energy models of correlated matter.

The Random Phase Approximation can be derived in several ways. In the original works~\cite{Bohm51,Bohm52,Bohm53}, the approximation was introduced to decouple momenta in the equation of motion.
A second point of view is as a self-consistent field approach to Coulomb screening~\cite{Ehrenreich59}. Finally, in a diagrammatic interpretation, the Random Phase Approximation corresponds to the summation of an infinite set of so-called link chain diagrams. Gell-Mann and Brueckner showed~\cite{Gell-Mann57} that this series is dominant in the electron gas at high density, thereby providing a solid theoretical justification for the use of the RPA. 

This proof applies to the limit $r_s \rightarrow 0$ with the Wigner-Seitz radius $r_s$ being proportional to the typical electron distance~\footnote{To be more precise, $(4 \pi/3) r_s^3$ is the volume per electron. Our interest here is in scaling and we drop all numerical factors for simplicity.}. To obtain a dimensionless expression for this limit, the electronic length scale $r_s\equiv \ell$ should be compared with the Thomas-Fermi screening length $L=\sqrt{r_s a_B}$, where $a_B$ is the Bohr radius. The RPA limit $r_s\rightarrow 0$ should then be read as $\ell/L = \sqrt{r_s/a_B} \ll 1$, relating the length scales of electronic propagation and Coulomb interaction in the electron gas. 

However, modern applications of RPA are by no means restricted to the dense electron gas. 
Insulators, with no states at the Fermi level, are very clearly not dense electron gases, yet the Random Phase Approximation performs admireably~\cite{Shishkin07}.
An important application of the RPA to insulators is the usage of the constrained Random Phase Approximation~\cite{Aryasetiawan04} (cRPA) to calculate the effective interactions between correlated electrons in a ``low-energy'' (or ``target'') subspace of materials such as transition metal compounds~\cite{Miyake09,Sasioglu11,Vaugier12,Sakuma13,vanLoon18c}, graphene~\cite{Wehling11}, cuprate~\cite{Werner15,Jang16} and nickelate~\cite{Nomura19} superconductors and other materials~\cite{Martins11,Shih12}. 
The ``target'' space is subsequently treated with more accurate methods~\cite{Anisimov97,Lichtenstein98,Kotliar06} that are able to deal with strong correlations. 
In the cRPA, the partially screened interaction is calculated by excluding from the RPA diagrams all virtual excitations that occur entirely in the target space, i.e., close to the Fermi level.
Therefore, the properties of the cRPA in a system with a gapped rest space are very similar to those of RPA in a gapped system. In this way, the RPA plays a central role in the modern understanding of both semiconductors and strongly correlated materials.

This raises the question if and how the (c)RPA approximation for screening can actually be justified away from the dense electron gas limit of Gell-Mann and Brueckner, and in particular in systems with a gap. Particularly worrisome is the lack of electron-hole binding diagrams in the RPA, since this attractive interaction creates the excitons that are omnipresent in semiconductor physics~\cite{Mahan}. Can a theory that lacks these excitonic diagrams properly describe the dielectric properties of gapped materials?

Here, we show (1) that the distance from the Fermi level localizes the electronic propagator, (2) that the corresponding short electronic propagation length scale -- as compared to the interaction length scale -- eliminates non-local vertex corrections, and (3) that the underestimation of the true gap in the Kohn-Sham system~\cite{Sham83} compensates for the lack of local vertex corrections (in form of excitonic contributions) in the RPA. 
Altogether, this makes the (c)RPA series dominant in the wide band gap limit and when applied based on Kohn-Sham inputs. 
This explanation for the smallness of non-local vertex corrections to (c)RPA can be seen as a real space, electron-electron analog of Migdal's theorem~\cite{Migdal58,Schrieffer18,Roy14} on the smallness of vertex corrections in electron-phonon problems.  

The paper is structured as follows: In Sec.~\ref{sec:lengthscale}, we show that the distance to the Fermi level indeed localizes the electronic Green's function. In Sec.~\ref{sec:nonlocal}, we use this localization to show the smallness of non-local vertex corrections to the screening. Section~\ref{sec:local} shows the compensation between local vertex corrections and Kohn-Sham energies used in density functional theory. This completes our theoretical arguments. In Sec.~\ref{sec:materials}, we then illustrate the screening length scales in cRPA calculations of graphene and SrVO$_3$, using spatial fluctuation diagnostics, after discussing how our arguments translate to the orbital basis sets used in actual cRPA calculations. Additional details, examples and discussion are available in the Appendices.

\section{Electronic length scale}
\label{sec:lengthscale}

We start with the first point: the energy offset of the screening bands from the Fermi level leads to a short propagation length for low-energy excitations.
For a simple parabolic band $E(\kv)=E_0+\hbar^2 k^2/(2m)$, 
the offset $E_0$ acts as a quantum tunneling barrier, $m$ is the (effective) electron mass and $\abs{\kv}<\pi/a$, with $a$ the lattice constant. 
We study the Green's function $G(E,\kv)$ in the limit of large $E_0$, i.e., $E_0 \gg E$ and $E_0 \gg \frac{\hbar^2}{m a^2}$, and expand it in $(E-E_0)^{-1}$,
\begin{align}
 G(E,\kv) = \frac{1}{E-E_\kv} \approx \frac{1}{E-E_0} \left( 1 + \frac{1}{2} \frac{\hbar^2 k^2}{m (E-E_0)} + \ldots \right).
\end{align}
A Fourier transform to real space,  $G(\rv) = \frac{a}{2\pi}\int_{-\pi/a}^{\pi/a} d\kv \cos(\kv \cdot \rv) \, G(E,\kv)$, gives 
\begin{align}
 G(\rv=0) &\approx  \frac{1}{E-E_0} \left(1 + \frac{\pi^2}{6} 
 \frac{\hbar^2}{m (E-E_0) a^2} + \ldots \right), \notag \\
 G(\rv=na) &\approx  -\frac{\hbar^2}{m  (E-E_0)^2 a^2 n^2} + \ldots,\text{ for $n\geq 1$.}  
\end{align}
The nonlocal part of $G$ is smaller than the local part by a factor $\frac{\hbar^2}{m (E-E_0) a^2} \ll 1$.  Beyond nearest-neighbours, the Green's function decays with $\abs{\rv}$.~\footnote{Usually, for example also in the tight-binding model discussed below, the decay is exponential, as expected from Fourier theory. Here, we find only algebraic decay because the direct periodization of the function $k^2$ is not smooth at the edge of the Brillouin Zone, breaking the mathematical requirement for exponential decay of the Fourier transform. }
For holes, the same argumentation holds by taking both $E_0<0$ and $m<0$.

From a tight-binding perspective, the same short propagation length arises~\cite{Rosner16} from a small ratio of the hopping $t$ compared to the on-site potential $E_0$. As shown in Appendix~\ref{app:lengthscales}, in this situation the Green's function decays exponentially,
\begin{align}
G(E,\rv-\rv') &\approx \frac{1}{E-E_0} \exp\left(-\abs{\rv-\rv'}/\ell\right), 
\end{align}
with decay length $\ell$,
\begin{align}
 \ell &= \frac{a}{\ln(\abs{\frac{E-E_0}{t}})}. 
\end{align}

In both examples, the offset $E_0$ creates an energy barrier for the electrons that can only be traversed via quantum tunneling, which leads to short-ranged propagation.   
This short length scale of the Green's function provides a powerful handle on diagrammatic theories~\cite{Rubtsov09} as used in the following.

\section{Absence of non-local vertex corrections to the screening}
\label{sec:nonlocal}

Screening describes the reduction of the bare Coulomb interaction $V(\qv)$ between two charge carries in the presence of further carries, which is quantified by the dielectric function~\footnote{For the cRPA, we are interested in the effective dielectric function in the \emph{low-energy} target space. For the usual RPA, we also restrict ourselves to static streening. Here and in the following, we set $\omega=0$. The role of frequency is discussed in Appendix~\ref{app:frequency}.},\footnote{Quantifying the cRPA screening via an effective dielectric constant $\epsilon(\qv)$ is by itself already an approximation, since it restricts the interaction in the effective model to the two-particle level~\cite{Maier12,Honerkamp12}. Similar gap and length-scale based arguments can be applied to this approximation.} $\epsilon(\qv)$, with $\epsilon^{-1}(\qv) = 1+\chi(\qv) V(\qv)$, where $\chi(\qv)$ is the fully interacting (charge) susceptibility. Computational approaches need to find good approximations for $\chi$, or equivalently, for the irreducible part (polarization) $\Pi$ with $\chi(\qv) = \Pi(\qv) / [1 + V(\qv) \Pi(\qv)]$~\footnote{In our sign convention, $\chi>0$ and $\Pi>0$. Other conventions exist in the literature.}.

\begin{figure}
\includegraphics{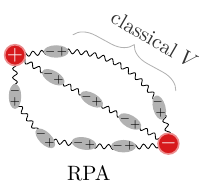}%
\quad\quad\quad%
\includegraphics{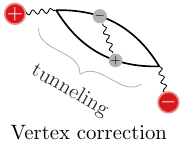}
\caption{Free charges (red) in an insulator are screened by particle-hole excitations (gray). The classical interaction $V$ allows screening over large distances (left) which is included in the RPA. On the other hand, the creation of particle-hole pairs (right) are classically forbidden and their propagation requires tunneling, restricting vertex corrections to short distances. In cRPA, the red charges denote the target band states and the gray excitations are from the rest space.}
\label{fig:rpasketch}
\end{figure}

\begin{figure}
\begin{align*}
\text{(a)} \quad
\chi^{\text{RPA1}}(\rv_2-\rv_1)=&
 \raisebox{-0.45\height}{\includegraphics{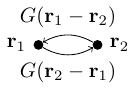}}
 &\substack{\#F=1\\\#C=1}
\\ 
\text{(b)} \quad
\chi^{\text{RPA2}}(\rv_2-\rv_1)=&
 \raisebox{-0.4\height}{\includegraphics{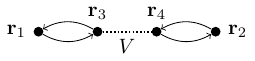}}
 & \substack{\#F=2\\\#C=2} 
\\ 
\text{(c)} \quad
 \chi^{\text{VX2}}(\rv_2-\rv_1)=&
 \raisebox{-0.45\height}{\includegraphics{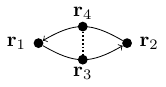}}%
 & \substack{\#F=1\\\#C=3} 
\end{align*}
\caption{Screening Feynman diagrams. Top: The lowest-order (Lindhard) screening process. Middle: The second order in (c)RPA. Bottom: A vertex correction not included in (c)RPA. 
On the right, the number of free ($\#F$) and constrained ($\#C$) spatial coordinates is shown, when considering that the Green's function (solid line) is localized.
}
\label{fig:diagrams}
\end{figure}

Several processes that contribute to $\chi$ are sketched in Fig.~\ref{fig:rpasketch}.
Fig.~\ref{fig:diagrams} shows the three lowest-order Feynman diagrams, where the dashed line stands for the bare interaction $V$ and the solid lines with arrows are electronic Green's functions $G$. The RPA corresponds to $\Pi=\chi^{0}$, where $\chi^{0}$ is the susceptibility of a non-interacting system. This creates a series of link-chain diagrams, with the lowest two orders illustrated in Fig.~\ref{fig:diagrams}(a) and (b). On the other hand, the vertex correction in Fig.~\ref{fig:diagrams}(c) is not included in the RPA. Here, we will show that the electronic propagation length scale $\ell$ controls the relative importance of diagrams (b) and (c). 

The real space coordinates $\rv_i$ involved in the screening processes are indicated in Fig.~\ref{fig:diagrams} as well. If the electrons are localized, then any two coordinates connected by an electronic Green's function (thick line) should be close together. For example, in diagram (a), after $\rv_1$ has been chosen freely, $\rv_2$ is constrained by the localization. In diagram (b), the localization of the Green's function requires that $\abs{\rv_1-\rv_3}$ and $\abs{\rv_2-\rv_4}$ are small, i.e., not exceeding the scale of $\ell$. There is no electronic constraint on $\abs{\rv_3-\rv_4}$. Instead, the Coulomb interaction sets this length scale, which is long-ranged (as discussed in Appendix~\ref{app:dipolar}, the relevant Coulomb matrix elements are dipolar).

The limit of electronic localization can be made more formal and precise, as is done in Appendix~\ref{app:explicitdiagrams}. For a brief summary, we consider a Green's function of the form $G(\rv_2-\rv_1) = G_0 \theta(\ell-\abs{\rv_2-\rv_1})$, where $\theta$ is the Heaviside step function. This constrains any integral over $\rv_2$ to a sphere with volume $4\pi\ell^3/3$ around $\rv_1$, i.e., the result of such an integral is proportional to $\ell^3$ in three dimensions. In this way, diagram (a) involves a single spatial constraint ($\#C=1$) and a factor $\ell^3$ and diagram (b) has two constraints ($\#C=2$) with a corresponding factor $\ell^6$.

The spatial constraints are essentially different for the corresponding vertex correction diagram in Fig.~\ref{fig:diagrams}(c), since all four coordinates are connected by electronic Green's functions, such that $\abs{\rv_4-\rv_1}<\ell$, $\abs{\rv_3-\rv_1}<\ell$ and $\abs{\rv_2-\rv_3}<\ell$. With these three constraints, $\#C=3$, the triangle inequality guarantees $\abs{\rv_2-\rv_4}<3\ell$, so the final constraint $\abs{\rv_2-\rv_4}$ does not provide an additional power of $\ell$ and only contributes a factor of order unity. 
Altogether, the three constrained variables suggest an overall power $\ell^9$. In fact, due to the $1/\abs{\rv_4-\rv_3}$ magnitude of the Coulomb interaction, the exponent is lowered to $\ell^8$ (see Appendix~\ref{app:explicitdiagrams}). Essentially, the electronic constraints keep the electron and hole close together, this makes the average magnitude of the Coulomb interaction larger than if all spatial integrals were entirely free. This effect lowers the exponent of $\ell$ by one. Comparing the powers of $\ell$ of the two diagrams, we come to one of our main results:
\begin{align}
\lim_{\ell\rightarrow 0}
 \underbrace{\chi(\qv)^{\text{VX2}} / \chi(\qv)^{\text{RPA2}}}_{\propto \ell^2} = 0.
\end{align}
In the limit of short electronic propagation lengths, i.e. for wide-gapped semiconductors, the RPA series is hence dominant over non-local vertex corrections.

The distinct length scales of vertex correction and RPA diagrams originate from the different roles of quantum and classical physics: in RPA only the creation of dipole moments as electron-hole pairs is a quantum process while the long-ranged screening results from the \emph{classical} electromagnetic (dipolar) interaction between these quantum fluctuations. On the other hand, the length scale of the vertex corrections is set by \emph{quantum tunneling} of electrons and holes (see Sec.~\ref{sec:lengthscale}), a process that is classically forbidden by the band gap and the resulting screening processes are thereby strongly localized. 

The smallness of non-local vertex corrections in RPA is reminiscent of Migdal's theorem for electron-phonon interactions, since both arguments are based on the phase space available for internal cooordinates in Feynman diagrams. Migdal's theorem is based on a phase space analysis of the \emph{momentum} space integrals in Feynman diagrams, which shows that vertex corrections are small compared to the geometric series of self-energy insertions.
The present argument considers \emph{real} space integrals and compares vertex corrections to the RPA geometric series of Lindhard bubble insertions. This analogy is discussed in more detail in Appendix~\ref{app:migdal}.

\section{Local vertex corrections}
\label{sec:local}

\subsection{Bonding-antibonding model} 
\label{sec:BABmodel}

Having shown that the polarization $\Pi$ is entirely local in systems with a (wide) gap, the remaining question is if the neglect of \emph{local} vertex corrections in (c)RPA is problematic.
To answer this, we utilize a minimal local model of an insulator consisting of two states with a bonding ($b$) and an antibonding ($a$) orbital at half filling and with a Hamiltonian in the single-particle eigenbasis of the form
\begin{align}
\label{eq:ed}
 H = 
 &\sum_{\sigma} 
 E_a n_{a,\sigma}
 +E_b n_{b,\sigma} \\
 +\frac{1}{2} &\sum_{\sigma,\sigma'} \sum_{\alpha\beta\gamma\delta} \Gamma_{\alpha\beta\gamma\delta} 
 f^\dagger_{\alpha\sigma} f^\dagger_{\gamma\sigma'} f^{\phantom{\dagger}}_{\delta\sigma'} f^{\phantom{\dagger}}_{\beta\sigma}.  \notag
\end{align}
Here, $\sigma,\sigma'\in\{\up,\dn\}$ and $\alpha\beta\gamma\delta\in \{a,b\}$ and $f_{a(b)}^\dagger$ and $f_{a(b)}$ denote the creation and annihilation operators for the antibonding (bonding) orbital. 
The respective density operators are $n_{a(b),\sigma}=f^\dagger_{a(b)\sigma} f^{\phantom{\dagger}}_{a(b)\sigma}$. 
$E_a$ and $E_b$ are the single-particle energies of the antibonding and bonding orbitals, respectively.
$\Gamma$ is the Coulomb interaction between the orbitals. In a situation without external screening it is given by the Coulomb integral
\begin{align}
 \Gamma_{\alpha\beta\gamma\delta} \approx \int dr \int dr' \phi_\alpha^\ast(r) \phi_\beta(r) V(r-r') \phi_\gamma^\ast(r') \phi_\delta(r'),
\end{align}
where $\phi_\alpha(r)$ is the wavefunction of the orbital $\alpha$ and $V(r-r')$ is the Coulomb interaction. To provide a representative example, we will use parameters derived from first principles that correspond to a single $sp^2$ bond in graphene discussed in more detail in Sec.~\ref{sec:graphene} and App.~\ref{app:vasp}. This model describes the screening by the graphene $sp^2$ orbitals in a constrained fashion, excluding any screening that results from the low-energy $p_z$ space. 

The two relevant single-particle Kohn-Sham energies are $h_b=-13.26\,$eV and $h_a=+12.58\,$eV, resulting in a Kohn-Sham gap $E_\text{KS}=25.84\,$eV. 
The cRPA~\footnote{Here, cRPA refers to taking both the $p_z$ and the $sp^2$ orbitals as the target space. This is discussed in more detail in Sec.~\ref{sec:materials}.} Coulomb interaction matrix elements are
$\Gamma_{aaaa}\equiv U_a=11.21\,$eV, $\Gamma_{bbbb}\equiv U_b=14.01\,$eV, $\Gamma_{aabb}\equiv V_{ab}=11.85\,$eV, $\Gamma_{abba}=\Gamma_{abab}\equiv J_{ab}=2.51\,$eV and all other non-zero Coulomb elements follow by symmetry.
Interactions with an odd number of $a$ or $b$ labels are forbidden because of the mirror symmetry of the orbitals. 
Only $J_{ab}$ leads to off-diagonal elements in the Hamiltonian. Since $J_{ab}$ is much smaller than the Kohn-Sham gap and the other interactions, it is justified to treat the states of Fig.~\ref{fig:minimalmodel} as approximate eigenstates.
This makes the analysis especially simple: we only need to calculate the energy differences between these states.

\newcommand{\cOne}{\text{\ding{192}}}
\newcommand{\cTwo}{\text{\ding{193}}}
\newcommand{\cThree}{\text{\ding{194}}}
\newcommand{\cFour}{\text{\ding{195}}}

In the ground state (\cOne), the bonding orbital is completely filled and the antibonding orbital is empty. The gap $E_g$ (also called true gap or electronic quasiparticle gap) of the system is defined by the difference between electron ionization and affinity energy, i.e.
\begin{align}
    E_g &= (E_{N-1} - E_{N}) - (E_N - E_{N+1})\\
    &= (E_\cTwo - E_\cOne)  - (E_\cOne - E_\cThree) \notag \\
    &= E_\cTwo + E_\cThree - 2 E_\cOne. \notag
\end{align}
At the bottom of Fig.~\ref{fig:minimalmodel}, the energies of these valence and conduction band states are denoted as $E_\text{VB}=E_\cOne-E_\cTwo$ and $E_\text{CB}=E_\cThree-E_\cOne$, respectively. 

The cost of a particle-hole excitation is substantially smaller than the difference of the valence and conduction energies due to the exciton binding, i.e. due to the attractive Coulomb interaction between the electron and the hole. We find
\begin{align}
    E_\cFour - E_\cOne = E_\text{CB}-E_\text{VB} - V_{ab} +J_{ab} < E_\text{CB} - E_\text{VB}.
\end{align}
With the numbers given above, exciton binding reduces the energy cost of this excitation by almost 10 eV. This suggests that the RPA, which does not include exciton binding diagrams, could struggle to properly describe screening in this model.

\subsection{True gap, Kohn-Sham gap and exciton binding}

\begin{figure}
\includegraphics{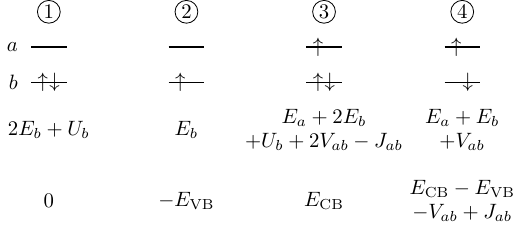}
\caption{Four relevant configurations in the minimal bonding-antibonding model with their corresponding energies $E_\cOne$, $E_\cTwo$, $E_\cThree$ and $E_\cFour$, in two different notations. Note that these configurations are not exact eigenstates of the interacting Hamiltonian, Eq.~\eqref{eq:ed}, the shown energies are the diagonal elements of the Hamiltonian.}
\label{fig:minimalmodel}
\end{figure}

Above, we discussed the difference between the true gap of the system, as measured by (inverse) photoemission, and the energy of particle-hole excitations, measurable in optical experiments. The Coulomb interaction, in particular the exciton binding, is responsible for their difference. To connect an \emph{ab initio} calculation to our bonding-antibonding model, we need to understand how to derive the single-particle model parameters $E_a$, $E_b$ from the Kohn-Sham energies $h_a$, $h_b$ that come from the Density Functional Theory (DFT) calculation. Formally, the Kohn-Sham energies are auxiliary quantities without direct physical meaning. In fact, it is well known that the Kohn-Sham gap $E_\text{KS}$ differs~\cite{Sham83} from the true gap $E_g$ by the derivative discontinuity $\Delta$, 
\begin{align}
E_g=E_\text{KS}+\Delta.
\end{align}
In our simple model, both gaps and the derivative discontinuity can be calculated. 
In an auxiliary non-interacting Kohn-Sham system we necessarily have $E^\text{KS}_\cTwo+E^\text{KS}_\cThree=3\epsilon^\text{KS}_b+\epsilon^\text{KS}_a=E^\text{KS}_\cOne+E^\text{KS}_\cFour$, since the total orbital occupations on both sides are the same.
In reality, however, we need to take correlation effects into account yielding a finite $(E_\cTwo+E_\cThree)-(E_\cOne+E_\cFour) = V_{ab}-J_{ab}>0$, which we immediately recognize as the exciton binding energy and which we identify below as the derivative discontinuity.
It accounts for the energy difference between a particle-hole excitation and the independent removal and addition of electrons.

\subsubsection{Determining the derivative discontinuity}

Establishing the derivative discontinuity $\Delta$ of the bonding-antibonding model is a central result of this work. It can be derived exactly by considering the average energy of an ensemble of realizations~\cite{Carrascal15} depicted in Fig.~\ref{fig:minimalmodel} and comparing this to the Kohn-Sham energies $\epsilon^\text{KS}$, which are constructed from a single DFT calculation at fixed integer density, in this case at $N=2$. The average energy is a functional of the occupations $n_a$ and $n_b$ and is given by
\begin{align}
    E(n_a,n_b) = n_\cOne E_\cOne+n_\cTwo E_\cTwo+n_\cThree E_\cThree + n_\cFour E_\cFour,
\end{align}
where the average occupations $n_i$ on the right-hand side are functions of $n_a$ and $n_b$ with the constraints $n_a=n_\cThree+n_\cFour$, $n_b=2n_\cOne+n_\cTwo+2n_\cThree+n_\cFour$, $N=n_a+n_b$. The Kohn-Sham energies are defined by $\epsilon^\text{KS}_a=\partial E/\partial n_a$ and $\epsilon^\text{KS}_b=\partial E/\partial n_b$, where the derivatives are evaluated at the ground state densities. Since we are interested in the energy functional close to this filling, we will write $N=2+\delta$ for the average number of electrons in the ensemble. 

\subsubsection{$\epsilon_b$}

For $\delta<0$ ($N<2$) the ground state is given by
\begin{align}
n_\cOne&=1+\delta,&  
n_\cTwo&=-\delta,&  
n_\cThree&=0,&  
n_\cFour&=0, \notag\\
n_b&=2+\delta,&  
n_a&=0. 
\notag
\end{align}
This direcly allows us to obtain
\begin{align}
    \epsilon_b &=\partial E/\partial n_b=d E/d \delta = E_\cOne-E_\cTwo=E_\text{VB}. \notag
\end{align}
The equality $\epsilon_b=E_{VB}$ is called the ionization potential theorem~\cite{Perdew97}. It states that the energy of the highest occupied Kohn-Sham level has physical meaning and is called the ionization potential.

\subsubsection{$\epsilon_a$}

To obtain $\epsilon_a$ as the derivative with respect to $n_a$, we need to consider the energy functional to linear order in $n_a$. Here, it will turn out that the sign of $\delta$ becomes important, in other words, if we are below or above $N=2$. For $N<2$ ($\delta<0$), the ensemble will contain realizations of state $\cTwo$, so the cheapest way to fill orbital $a$ is to replace a single realization of $\cTwo$ by $\cFour$ in the ensemble, i.e.,
\begin{align}
n_\cOne&=1+\delta,&  
n_\cTwo&=-\delta-n_a,&  
n_\cThree&=0,&  
n_\cFour&=n_a, \notag\\
n_b&=2+\delta,&  
n_a&=n_a.   &\multispan4{\text{(for $\delta=N-2<0$)}}
\notag
\end{align}
This gives the energy functional
\begin{align}
    E(n_a,n_b) = (1+\delta)E_\cOne -(\delta+n_a) E_\cTwo + 0 \, E_\cThree + n_a E_\cFour \notag
\end{align}
and we obtain 
\begin{align}
    \epsilon_a^{\delta<0}=\partial E/\partial n_a=E_\cFour-E_\cTwo.  \label{eq:derivdisc:1}
\end{align}
We note that this is not equal to $E_{CB}$.

Similarly, for $\delta \geq 0$ ($N \geq 2$),
the ensemble consists entirely of realizations of $\cOne$ and $\cThree$. To determine $\epsilon_a$, we need to add particles to orbital $a$, which can be done by adding electrons to the system,
\begin{align}
    n_\cOne&=1-n_a,&  
    n_\cTwo&=0,&  
    n_\cThree&=n_a,&  
    n_\cFour&=0, \notag\\
    n_b&=2,&  
    n_a&=\delta. 
\notag
\end{align}
and we get 
\begin{align}
    \epsilon_a^{\delta>0} &=\partial E/\partial n_a=dE/d\delta=E_\cThree-E_\cOne=E_{CB}. 
    \label{eq:derivdisc:2}
\end{align}
Comparing Eqs.~\eqref{eq:derivdisc:1} and \eqref{eq:derivdisc:2}, we see that $\epsilon_a$ has a discontinuity
\begin{align}
    \Delta &= \epsilon_a^{\delta>0} - \epsilon_a^{\delta<0} = (E_\cThree - E_\cOne ) - (E_\cFour-E_\cTwo) \notag\\
    &= V_{ab}-J_{ab}.
\end{align}
This is the same result that was derived heuristically as the difference between the non-interacting and interacting system before. For an extended discussion of the derivative discontinuity in two-orbital systems, we refer the reader to Ref.~\cite{Carrascal15}. A finite temperature interpretation of these results is available in Appendix~\ref{app:temperature}.

This derivation of $\Delta$ allows us to determine the Kohn-Sham gap analytically, 
\begin{align}
 E_\text{KS} &= h_a - h_b= E_g - \Delta = E_{CB}-E_{VB}-V_{ab}+J_{ab} \notag \\
 &= (E_a - E_b) -(U_{b}-V_{ab}). \notag
\end{align}
So, given the Kohn-Sham energies $h_a$ and $h_b$ and the interaction $\Gamma$, both from the \emph{ab initio} calculation, we have to set $E_a = h_a - V_{ab}$ and $E_b =h_b - U_b$ to ensure that the Hamiltonian~\eqref{eq:ed} has the same Kohn-Sham gap as the corresponding DFT calculation. 
The true gap of Eq.~\eqref{eq:ed} is $E_g = h_a-h_b + V_{ab}-J_{ab}$. In other words, the true gap of the model is larger than the gap in the underlying DFT calculation and the difference is exactly the exciton binding energy. For the example of graphene $sp^2$ states, the relevant energies are sketched in Fig.~\ref{fig:graphene:ed}(a).

\subsection{Screening: Cancellation of local vertex corrections}
\begin{figure*}[t]
\includegraphics[]{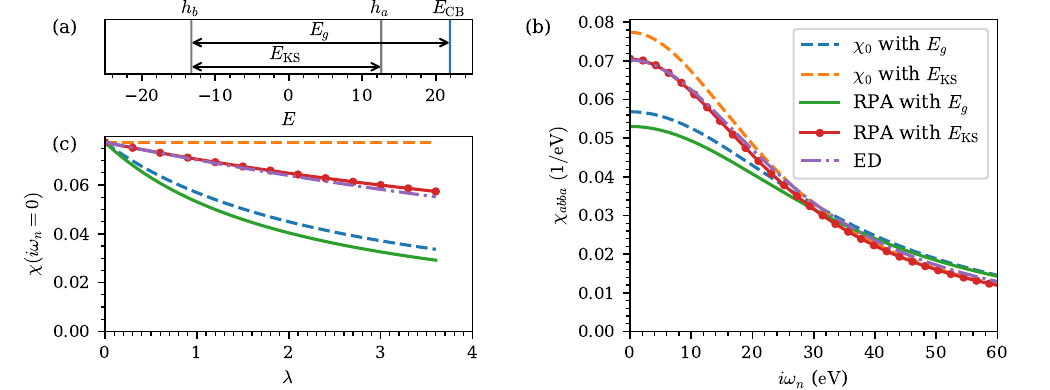}
\caption{Minimal bonding-antibonding model, Eq.~\eqref{eq:ed}. (a) Single-particle energies of the bonding-antibonding Hamiltonian. (b) Dynamical susceptibility on the Matsubara axis. Note that we show the charge susceptibility $\chi_{\up\up\up\up}+\chi_{\up\up\dn\dn}+\chi_{\dn\dn\up\up}+\chi_{\dn\dn\dn\dn}$.
(c) Susceptibilities with rescaled interaction $\Gamma\rightarrow \lambda \Gamma$, while keeping the Kohn-Sham gap $\epsilon_g$ constant.
}
\label{fig:graphene:ed}
\end{figure*}

After considering the effect of the Coulomb interaction on the single-particle properties of our \emph{ab initio} derived model Hamiltonian, we now proceed towards the screening properties as the central interest of this manuscript. Screening happens via particle-hole excitations that leave the total charge constant, so we should expect excitonic screening effects to be important and visible in the susceptibility of the system~\cite{Cunningham18}. 

The exact susceptibility of the bonding-antibonding Hamiltonian can be calculated using exact diagonalization~\cite{pomerol} (ED) which can be compared to approximations. We consider the charge susceptibility $\sum_{\sigma\sigma'}\chi_{\sigma\sigma\sigma'\sigma'}$ here and restrict ourselves to the particle-hole channel. In the interacting susceptibility, orbital combinations such as $\chi_{abab}$ or $\chi_{aabb}$ are also allowed but they remain an order of magnitude smaller, since $J_{ab}$ is small.

The blue dashed line in Fig.~\ref{fig:graphene:ed}~(b) is the non-interacting susceptibility using the \emph{true gap}, i.e., $\chi_0(i\omega_n)_{\sigma\sigma\sigma'\sigma'} = \delta_{\sigma\sigma'}\Re \,\, (E_{\text{CB}}-E_{\text{VB}}+i\omega_n)^{-1}$, while the orange dashed line corresponds to the non-interacting susceptibility using the Kohn-Sham gap. The purple dash-dotted line represents the exact susceptibility as derived from an exact diagonalization of our model Hamiltonian, which is between the two bare susceptibilities.

In addition to the non-interacting and the exact susceptibilities, we also show the interacting charge susceptibilities, $\chi_\text{RPA}(i\omega_n)_{\sigma\sigma\sigma'\sigma'} = 2\chi_0(i\omega_n) / \left[1 + J_{\text{ab}} \chi_0(i\omega_n)\right]$ as obtained within RPA, with the factor 2 originating in the spin sum. We have neglected the orbital matrix structure of $\chi$ here, which leads to mixing of $\chi_{abba}$ and $\chi_{abab}$ in the RPA equations. Since $(\chi_0)_{abab}=0$, this effect is however negligible in the present case. The resulting RPA expression does not contain $V_{ab}$, which is responsible for the exciton binding, and therefore reduces the susceptibility. Starting from the true gap $\chi_0$, this makes the results worse (green solid curve). 
However, starting from the bare susceptibility calculated with the Kohn-Sham gap $E_\text{KS}$, $\chi_\text{RPA-KS}(i\omega_n)_{\sigma\sigma\sigma'\sigma'} = 2\chi_{0,\text{KS}}(i\omega_n) / \left[1 + J_{\text{ab}} \chi_{0,\text{KS}}(i\omega_n)\right]$ the ``miracle'' of RPA for insulators occurs: the red symbols are on top of the exact result. 
Apparently, a cancellation occurs between the underestimation of the true gap in the Kohn-Sham construction and the lack of local exciton vertex corrections in RPA due to the lack of $V_{ab}$. In other words: since the Kohn-Sham gap and the RPA susceptibility are both calculated within the charge-neutral system, the reduction of the Kohn-Sham gap due to excitonic screening renders the explicit diagrammatic  treatment of excitonic screening unnecessary when calculating the RPA susceptibility starting from Kohn-Sham states.

As stated earlier, the relatively simple picture of the bonding-antibonding model is applicable as long as the states of Fig.~\ref{fig:minimalmodel} are a good approximation of the eigenstates of Eq.~\ref{eq:ed} which holds if $J_\text{ab}$ is small. 
If we change the relative strength of all interactions by rescaling the Coulomb vertex, $\Gamma \rightarrow \lambda \Gamma$, while keeping the Kohn-Sham gap $E_\text{KS}$ fixed, the susceptibility changes, as shown in Fig.~\ref{fig:graphene:ed}~(c). 
While the non-interacting susceptibility based on the Kohn-Sham gap $E_\text{KS}$ does not depend on $\Gamma$ and thus also not on $\lambda$, the non-interacting susceptibility based on $E_g$ changes with $\lambda$ since $E_g$ is affected.
The RPA with $E_\text{KS}$ performs well for all shown values, although the deviations increase with $\lambda$. In the ED, this (slow) breakdown of the occupation number eigenbasis due to Coulomb interactions is also visible as a finite occupation $n_a$ of the antibonding orbital in the many-body ground state.

Thus, we have shown that RPA based on the Kohn-Sham energies gives an accurate description of the polarizability of bonding-antibonding states, as long as the gap is large enough that no appreciable changes in the occupation numbers $n_a$ and $n_b$ occur. The neglect of vertex corrections works so well because the Kohn-Sham gap already contains the leading vertex correction, namely the exciton binding energy.

Finally, we note that the cancellation of vertex corrections to $\chi$ in this simple model implies the same for the local vertex corrections to $\Pi$, since $\Pi$ is just the irreducible part of $\chi$. This completes the proof of the applicability of (c)RPA for systems with a wide gap.

\section{Screening and cRPA effective interactions in materials}
\label{sec:materials}

After these general considerations on electronic screening in insulators, it is useful to study some examples in detail. Here, we will investigate the cRPA determination of effective Hubbard interactions in graphene, where we focus on the screening by $sp^2$ states, and in SrVO$_3$, where we study the screening by O $p$ and V $e_g$ states. All \emph{ab initio} calculations are performed using VASP and all technical details can be found in Appendix~\ref{app:vasp}.
We use TRIQS~\cite{Triqs2015} and tprf~\cite{tprf} to further analyze and manipulate the resulting quantities.

\subsection{Orbital structure of RPA}

Unlike in the homogeneous electron gas, in real materials we have to take into account the orbital or band structure of the electrons in our calculations. 
In that case, the bare and screened interactions are related by the self-consistent relation
\begin{align}
 U_{ab,cd} &= V_{ab,cd} - V_{ab,fe} \Pi_{ef,gh} U_{hg,cd} \notag \\
 &=V_{ab,cd} - U_{ab,fe} \Pi_{ef,gh} V_{hg,cd},  \label{eq:bse}
 \end{align}
which is called the two-particle Dyson or Bethe-Salpeter equation~\cite{Nakanisha69}. Here, the letters are the combined electronic orbital and spatial indices and summation over internal orbital labels is implied.
Diagrammatically the equation can be visualized as
\includegraphics{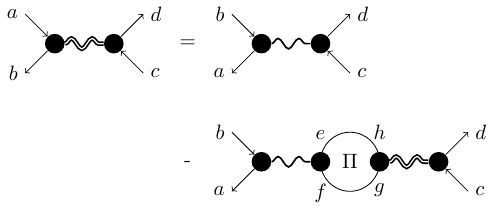}
where the double (single) snaked line stands for the dressed interaction $U$ (bare interaction $V$).

\subsection{Multi-tiered RPA}

\begin{figure}
    \centering
    \includegraphics{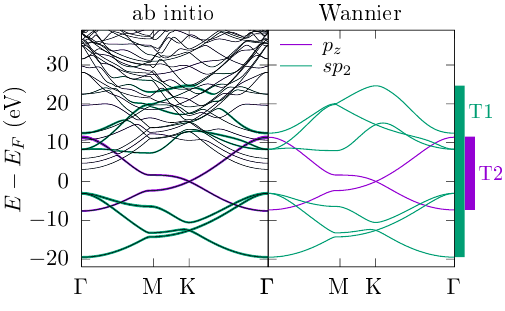}
    \caption{Ab initio (left) and Wannierized (right) band structure of graphene. The $T2$ subspace consists of the bands with $p_z$ character (purple), the $T1$ subspace also includes those with $sp_2$ character (green). }
    \label{fig:graphene}
\end{figure}

The evaluation of the RPA screening can be done in several steps, generating an intermediate partially dressed interaction by integrating out some electronic states and then using this as the ``bare'' interaction for the next downfolding step. This multi-tiered approach~\footnote{For another multi-tiered approach, see Ref.~\cite{Nilsson17}.} is illustrated in Fig.~\ref{fig:graphene}, where $T1$ (all $sp^3$ states) and $T2$ (just $p_z$ states) are two subspace of the full electronic structure $\Omega$ with $T2 \subset T1$. 

The partially screened interaction of a specific tier is calculated by including all screening processes except those that occur entirely within that tier. Explicitly we can define $U^{T1}$ and $U^{T2}$ via
\begin{align}
 \left(U^{T1}\right)^{-1} &= \left(V\right)^{-1} + \Pi^{T1} \\
 \left(U^{T2}\right)^{-1} &= \left(V\right)^{-1} + \Pi^{T2} \label{eq:tier12} \\
                          &= \left(U^{T1}\right)^{-1} + \underbrace{(\Pi^{T2}-\Pi^{T1})}_{\equiv \Pi^{T1\rightarrow T2}} \notag
\end{align}  
The involved polarizations are calculated as the difference between the polarization of the full electronic structure and the polarization that is entirely within the tier,
\begin{align}
    \Pi^{T2} &= \sum_{\alpha\beta\gamma\delta\in \Omega} \Pi_{\alpha\beta\gamma\delta} - 
                \sum_{\alpha\beta\gamma\delta\in T2} \Pi_{\alpha\beta\gamma\delta}, \\
    \Pi^{T1} &= \sum_{\alpha\beta\gamma\delta\in \Omega} \Pi_{\alpha\beta\gamma\delta}- 
                \sum_{\alpha\beta\gamma\delta\in T1} \Pi_{\alpha\beta\gamma\delta},
\end{align}
where in RPA the polarization operator tensor elements are defined by
\begin{align}
    \Pi_{\alpha\beta\gamma\delta} = -G_{\alpha\delta}G_{\beta\gamma}.
\end{align}
Note that the constrained RPA is usually constructed in this way~\cite{Kaltak}, by selecting the target space as $T2$ and setting $T1 = \Omega$. Combining the equations, we find the relation
\begin{align}
    \Pi^{T1\rightarrow T2} = \sum_{\alpha\beta\gamma\delta\in T1} \Pi_{\alpha\beta\gamma\delta} -
                             \sum_{\alpha\beta\gamma\delta\in T2} \Pi_{\alpha\beta\gamma\delta}.
                             \label{eq:pi:diff}
\end{align} 
In other words, this polarization includes all excitations that occur within $T1$ except the ones that fall entirely within the target space $T2$. In terms of the band basis of Fig.~\ref{fig:graphene}, this corresponds to the following diagrams:
\begin{align}
\Pi^{T1\rightarrow T2} =
\begin{tikzpicture}[baseline=-5pt]
\draw[thick,ourGreen,bend left] (1.,0) to node {$<$} node[midway,below] {\small{T1}} (0,0) ;
\draw[thick,ourGreen,bend left] (0,0) to node {$>$} node[midway,above] {\small{T1}} (1.,0) ;
\end{tikzpicture}+
\begin{tikzpicture}[baseline=-5pt]
\draw[thick,ourGreen,bend left] (1.,0) to node {$<$} node[midway,below] {\small{T1}} (0,0) ;
\draw[thick,ourPurple,bend left] (0,0) to node {$>$} node[midway,above] {\small{T2}} (1.,0) ;
\end{tikzpicture}+
\begin{tikzpicture}[baseline=-5pt]
\draw[thick,ourPurple,bend left] (1.,0) to node {$<$} node[midway,below] {\small{T2}} (0,0) ;
\draw[thick,ourGreen,bend left] (0,0) to node {$>$} node[midway,above] {\small{T1}} (1.,0) ;
\end{tikzpicture}.
\end{align}
The latter two diagrams are usually called ``mixed'' diagrams. In the first diagram, both the electron and the hole are far away from the Fermi level, whereas in the latter two diagrams this only holds for one of them.

We will use this kind of two-step downfolding to study in detail how specific bands contribute to the screening and thus to get a quantitative understanding of all (c)RPA screening processes. We will do this by explicitly calculating $U^{T1}$ and $U^{T2}$ as tensors in orbital space from first principles. We also calculate $\Pi^{T1\rightarrow T2}$ from the band structure. The inversion of Eq.~(\ref{eq:tier12}) yields.
\begin{align}
    U^{T2} = U^{T1} - U^{T1} \Pi^{T1\rightarrow T2} U^{T2}. \label{eq:bse:2}
\end{align}
Since we have access to all quantities on both sides of the equation, we can easily evaluate the contributions of various screening processes.
This so-called fluctuation diagnostics~\cite{Gunnarsson15} is an efficient tool to identify which components of $\Pi$ are most responsible for screening.

\subsection{Orbitals, unit cells, tensors, and basis sets}

Before moving on to specific materials, we need to consider another aspect of the many-body theory of materials. 
We will consider crystalline materials with a Bravais lattice $\{\Rv_j\}$, where capital $\Rv_j$ stand for discrete vectors in the Bravais lattice and small $\rv$ for continuous coordinates. 
The relevant electronic spaces are spanned by a set of orbitals $\phi_{a,\Rv_j}$ centered on unit cell $\Rv_j$, where $a=1,\ldots N$ is the orbital label and $N$ the dimension of this electronic space.

Although the orbital $\phi_{a,\Rv_j}$ is centered on unit cell $\Rv_j$, it is not guaranteed that this orbital is entirely contained within the unit cell, in the sense that $\phi(\rv)=0$ must not hold for $\rv$ outside of the unit cell. For example, a Wannier construction will usually~\cite{Brouder07} yield orbitals $\phi_{b,\Rv_i}$ which are exponentially decaying, i.e. $\phi_{b,\Rv_i}(\rv)\sim \exp(-\kappa\abs{\rv-R_i}) > 0 $ for some $\kappa>0$ and large $\abs{\rv-R_i}$. 

The susceptibility, polarization and Coulomb interaction are two-particle Green's functions, so they have four orbital labels and must transform as rank-4 tensors~\cite{Kaltak} in orbital space. 
In the same vein, two-particle Green's functions generally involve four spatial coordinates, or three momenta.
In the electron gas, the Coulomb interaction is responsible for eliminating two coordinates/momenta: the Coulomb interaction $V(\rv_1,\rv_2,\rv_3,\rv_4)=V(\abs{\rv_1-\rv_4})\delta_{\rv_1,\rv_2}\delta_{\rv_3,\rv_4}$ acts not between four field operators at $\rv_1$, $\rv_2$, $\rv_3$, $\rv_4$ but between two densities $n(\rv_1)$ and $n(\rv_4)$.

This is no longer true when we transform the field operators to a combined cell/orbital basis if these orbitals are not entirely localized within the (unit) cell, as it is the case for realistic Wannier constructions. 
Thus, after the transformation to Wannier orbitals, the Coulomb interaction has four spatial indices $V(\Rv_1,\Rv_2,\Rv_3,\Rv_4)$, or three momenta $V(\Qv,\Kv,\Kv')$. We again use capitals (e.g., $\Qv$) to denote the Brillouin Zone momentum corresponding to the discrete vectors $\Rv$.

Since objects with three momenta are computationally inconvenient we will use below the approximation $V(\Qv,\Kv,\Kv')\approx V(\Qv)=\int d\Kv \int d\Kv' V(\Qv,\Kv,\Kv')$, where a suitable normalization of the integrals is implied. In real space, this corresponds to $V(\Rv_1,\Rv_2,\Rv_3,\Rv_4)=V(\Rv_1-\Rv_4)\delta_{\Rv_1,\Rv_2}\delta_{\Rv_3,\Rv_4}$, i.e., we assume that the pair of orbital operators on either side of the Coulomb interaction line shares a unit cell~\footnote{Instead of completely ignoring the dependence on $\Kv$ and $\Kv'$, one could also consider a form factor expansion~\cite{Husemann09}, but this is beyond the scope of this work.}. In making this approximation, we have to consider that the construction of the orbitals is not unique in several ways (choice of unit cell, choice of the orbitals, placement of the orbitals within the unit cell), as will be discussed in the case of graphene. The resulting spatial-orbital structure of $V$, expressed in terms of discrete unit-cell and orbital based coordinates, therefore explictly depends on the chosen orbital parametrization. The quality of the approximation $V(\Qv,\Kv,\Kv') \approx V(\Qv)$ can only be quantified within the context of a specific orbital basis.
As this issue is related to spillage of the orbitals outside of the unit cell, the approximation will perform well for sets of orbitals that are sufficiently localized within the unit cell. 
In Appendix~\ref{app:benzene}, we consider the benzene molecule where the computational cell effects can be seen quantitatively. 

 \subsection{Orbital basis for graphene}

 \label{sec:graphene}
 
 \begin{figure}
  \includegraphics[width=\columnwidth]{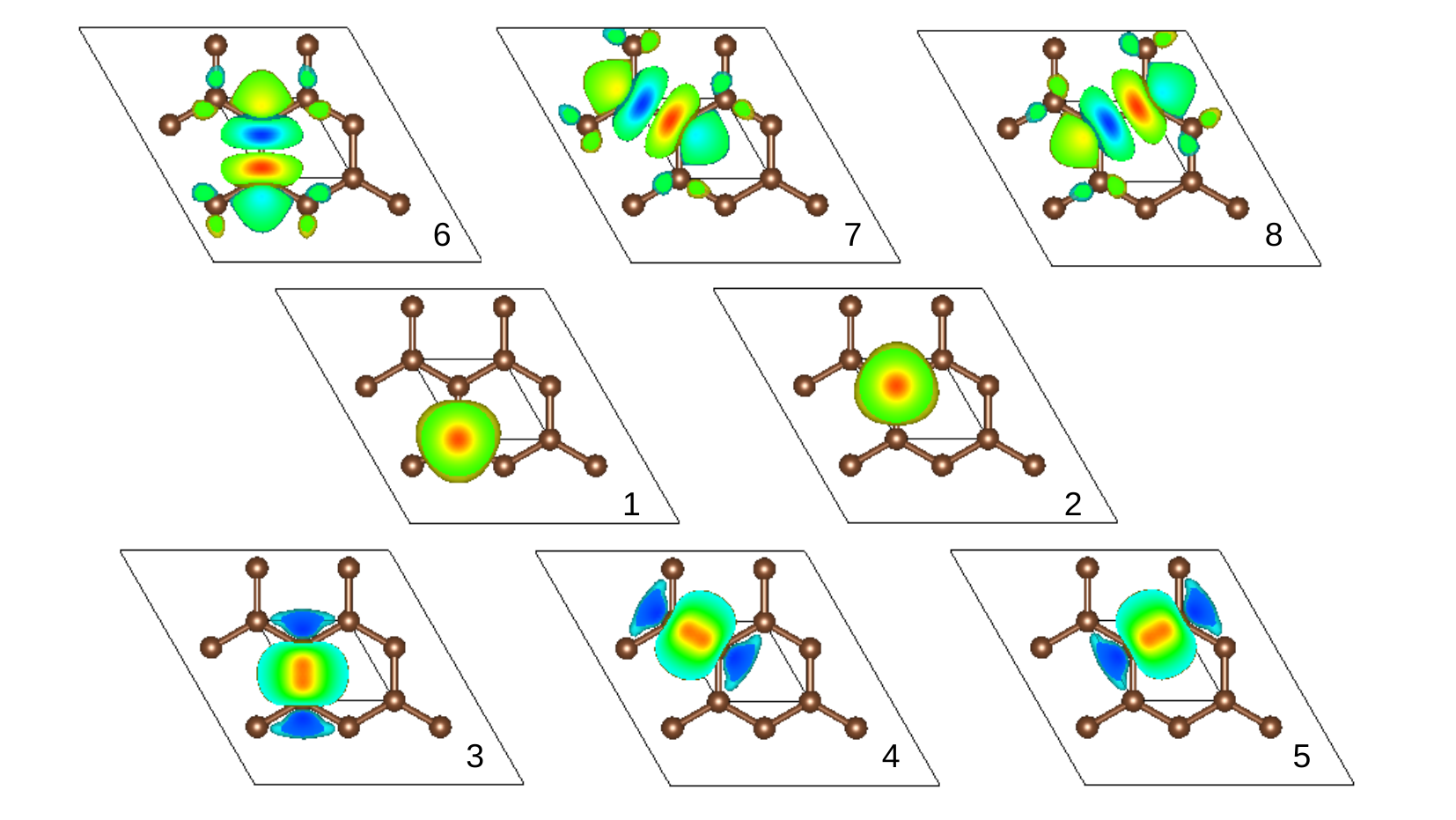}
  \caption{Wannier orbital bonding-antibonding basis for the $sp^2$ and $p_z$ states in graphene. Orbitals 1 and 2 are $p_z$ orbitals. Orbitals 3, 4 and 5 are $sp^2$ bonding orbitals and orbitals 6,7 and 8 are $sp^2$ anti-bonding orbitals.}
  \label{fig:graphene:wannier}
 \end{figure}
 
Graphene has two C atoms per unit cell and the lower-energy electronic structure is determined by their 2s and 2p electrons (the 1s states are far away). The two $p_z$-like orbitals generate two Dirac bands that linearly cross the Fermi level, as depicted in the band structure of Fig.~\ref{fig:graphene}. The corresponding real-space Wannier functions are shown in Fig.~\ref{fig:graphene:wannier} (1,2). The $sp^2$ orbitals can be classified as fully occupied bonding (3, 4, 5) and completely empty antibonding (6, 7, 8) orbitals further away from the Fermi level \cite{Marzari12}. Their electronic dispersions are shown in Fig.~\ref{fig:graphene} and their real-space form is shown in Fig.~\ref{fig:graphene:wannier}. As indicated in Fig.~\ref{fig:graphene}, we define the $T1$ space to consist of all eight orbitals and the $T2$ space contains just the two $p_z$-like orbitals. Our cRPA evaluation of the Coulomb matrix elements are performed using these orbitals. 

According to the definition of $\Pi^{T1\rightarrow T2}$ in Eq.~\eqref{eq:pi:diff}, the multi-tiered approach involves so-called ``mixed'' diagrams, where some but not all of the labels $\alpha\beta\gamma\delta$ lie in $T1$. In the present case, these diagrams do not contribute due to symmetry. 
The $p_z$ orbitals are antisymmetric under mirror symmetry in the plane, whereas the $sp^2$ orbitals are symmetric. Selection rules arise from this difference in symmetry and all quantities with an odd-number of $p_z$ labels vanish. In particular, this holds for Coulomb matrix elements such as $V_{p_z p_z p_z sp^2}=0$. This is the interaction that would couple to the ``mixed'' $\Pi$ in Eq.~\eqref{eq:bse:2}. As a result, for multi-tiered cRPA in graphene, we only need to consider the polarization $\Pi^{T1\rightarrow T2}_{abcd}$ for $a,b,c,d\in sp^2$.

As discussed above, the labels $a,b,c,d$ in Eq.~\eqref{eq:bse:2} should be combined orbital and unit cell labels, in the sense that $\Rv_a\neq \Rv_b$ is allowed if orbitals in different unit cells overlap. However, in the repeated downfolding, we neglect this additional dependence, i.e., we assume $V(\Qv,\Kv,\Kv')\rightarrow V(\Qv)$ and the labels $a,b,c,d$ are now only orbital labels. This is an approximation that will lead to deviations between the two-tiered approach and the direct cRPA downfolding. As a result of the approximation, we consider all objects as $8\times 8\times 8 \times 8$ orbital tensors with an additional $\Qv$ dependence on an $18\times 18$ grid, which combines to approximately 1.3 million matrix elements each for the objects $U$, $V$ and $\Pi$.

We should note that the placement of orbitals in the unit cell (Fig.~\ref{fig:graphene:wannier}) breaks the sublattice symmetry, in the sense that orbital 2 is surrounded by three bonds \emph{within} the unit cell whereas two of the bonds surrounding orbital 1 lie in neighbouring unit cells. Together with the simplification $V(\Qv,\Kv,\Kv')\rightarrow V(\Qv)$ (i.e., both orbitals on either end of the Coulomb interaction have to share a unit cell), this results in different on-site interactions on the two $p_z$ orbitals in the multi-tiered cRPA. We found that the deviation with the direct cRPA calculation is smallest for orbital 1 and therefore consider the on-site $p_z$ interaction on this orbital in the following. 

\subsection{Spatial fluctuation diagnostics for graphene}

\begin{figure*}
    \includegraphics{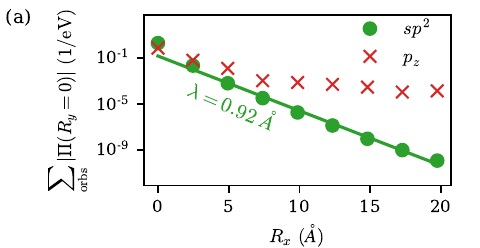}%
    \includegraphics{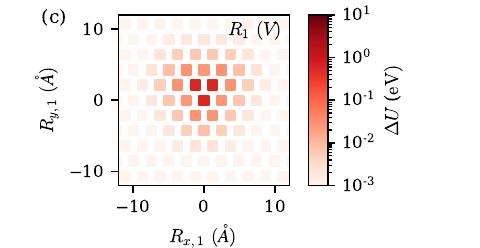}%
    \\
    \includegraphics{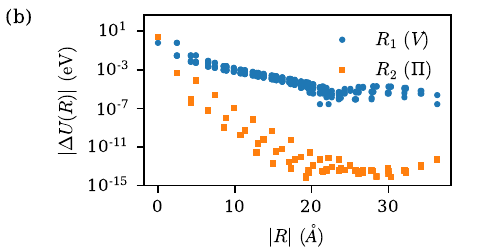}%
    \includegraphics{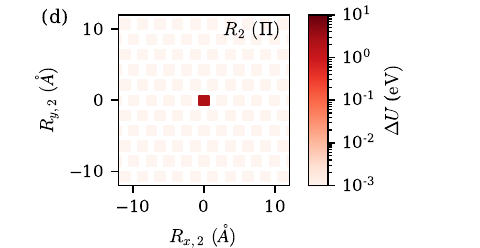}
    \caption{Spatial analysis of screening in graphene. (a) The $sp^2$ polarization is exponentially localized, the $p_z$ polarization is not. (b) Spatial fluctuation diagnostics in graphene according to Eqs.~\eqref{eq:flct:pi} and \eqref{eq:flct:v}. These are the contributions to the fully screened on-site $p_z$ interaction. In total, in the $T1\rightarrow T2$ downfolding, the $sp^2$ states screen the interaction down from 
    $U^{T1}=12.59\,$eV to $U^{T2,\text{indirect}}=10.25\,$eV ($\Delta U=2.70\,$eV). 
    (c,d) Fluctuation diagnostics in real space. The color of every square shows the magnitude of the screening contribution $\Delta U$ coming from a specific unit cell displacement $\Rv$.  }
 \label{fig:graphene:flct:r}
\end{figure*}

The exponential electronic localization of the rest space is directly visible in $\Pi^{sp^2}(\Rv)$. Figure~\ref{fig:graphene:flct:r}(a) shows the sum of absolute values of all orbital elements of $\Pi^{sp^2}$. The exponential decay occurs with a decay length of less than 1 \AA. This scale is well approximated by the expression (Appendix~\ref{app:lengthscales}) $a /\ln(E_0^2/t^2)$ with $E_0\approx 12\,$eV the distance of the $sp^2$ bands to the Fermi level (half the gap) and $t\approx 3\,$eV the hopping in the $sp^2$ manifold. The polarization in the target $p_z$ (red crosses) space is notably different and does not decay exponentially. The $p_z$ bands cross the Fermi surface which leads to a long-ranged polarization.

For the \emph{fluctuation diagnostics} of Eq.~\eqref{eq:bse:2}, our interest is to find the typical length scales involved in screening. 
To this end we write Eq.~\eqref{eq:bse:2} as a convolution in real space and concentrate on the on-site interaction in the final down-folded model,
\begin{align}
 U^{T2}_{\Rv\!=\!0}  = U^{T1}_{\Rv=0}
                     - \sum_{\Rv_1,\Rv_2} \! U^{T1}_{\Rv_1} \, \Pi^{T1\rightarrow T2}_{\Rv_2}\, U^{T2}_{-\Rv_1-\Rv_2},
\end{align}
where all objects are tensors in orbital space. In terms of the notation of Eq.~\ref{eq:bse}, the labels $a$, $b$, $c$ and $d$ correspond to a single $p_z$ orbital at $\Rv=0$, $e$ and $f$ are $sp^2$ orbitals at $\Rv_1$, and $h$ and $g$ are $sp^2$ orbitals at $\Rv_2$.

In the following we perform the distance fluctuation diagnostics on $\Pi$ and on $V$ via
\begin{align}
  U^{T2}_{\Rv\!=\!0}  = U^{T1}_{\Rv=0} 
                      - \sum_{\Rv_2} \Delta U_{\Rv_2} = U^{T1}_{\Rv=0}  
                      - \sum_{\Rv_1} \Delta U_{\Rv_1} \notag
\end{align}
with 
\begin{align}
    \Delta U_{\Rv_1} &= \sum_{\Rv_2} U^{T1}_{\Rv_1} \, \Pi^{T1\rightarrow T2}_{\Rv_2} \, U^{T2}_{-\Rv_1-\Rv_2} \label{eq:flct:v}, \\
    \Delta U_{\Rv_2} &= \sum_{\Rv_1} U^{T1}_{\Rv_1} \, \Pi^{T1\rightarrow T2}_{\Rv_2} \, U^{T2}_{-\Rv_1-\Rv_2} \label{eq:flct:pi}.
\end{align}
Figure~\ref{fig:graphene:flct:r}~(b,c,d) show the results of these distance fluctuation diagnostics. The screening contribution decays exponentially as a function of the distance covered by the propagating electron-hole pair $\Rv_2$ [orange squares in (b); (d)] and all contributions except for $\Rv_2=0$ are negligible. Thus, the electron-hole excitation in the screening process is localized, i.e., it does not propagate. On the other hand, the decay with respect to the Coulomb distance $\Rv_1$ is relatively mild [blue circles in (b); (c)].
Hence, all necessary perquisites needed for cRPA evaluation of the Coulomb matrix elements to perform well based on Kohn-Sham band structures are fulfilled so that it can safely be applied here.

\subsection{SrVO$_3$}

\begin{figure}
 \includegraphics{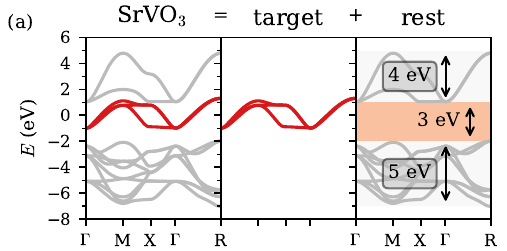}
 \includegraphics{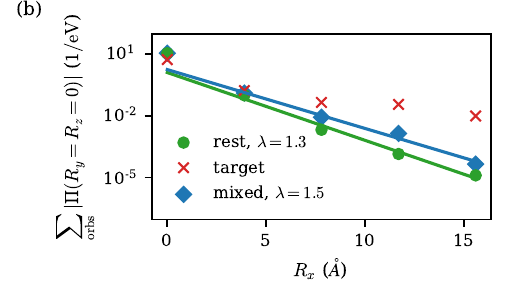}
 \caption{SrVO$_3$. (a) the band structure with target space (red) and rest space (grey). The rest space has a gap of 3 eV and this leads to an exponential localization of the rest space electrons, justifying the use of cRPA. (b) This exponential localization is visible in the rest and mixed polarization $\Pi$.}
 \label{fig:SrVO3} 
\end{figure}

SrVO$_3$ is a strongly correlated material for which Coulomb matrix elements are frequently evaluated from first principles using cRPA~\cite{Miyake08,Nomura12,Vaugier12,Taranto13,Kaltak}. So far the validity of the latter has not been discussed in detail.
Figure~\ref{fig:SrVO3}(a) shows the band structure including the relevant energy scales. We consider a target space consisting of three V $t_{2g}$ orbitals per unit cell, and a rest space of nine O $p$ and two V $e_g$ states per unit cell. 

The rest space has a 3 eV gap and as a result the rest space polarization $\Pi$ decays exponentially as a function of $\Rv$, Fig.~\ref{fig:SrVO3}(b), with a decay constant of $\lambda\approx 1.3\,$\AA, which is substantially smaller than the lattice constant $a\approx 3.9\,$\AA. The decay constant matches qualitatively with an analytical estimate based only on the energies of the bands, $\lambda \approx 0.84\,$\AA, see Appendix~\ref{app:lengthscales}. 

Unlike in graphene, in SrVO$_3$ the so-called ``mixed'' polarization, with one rest and one target space Green's function, also plays a role in the screening.
The localization argument applies only to the rest space Green's function. As a result, the mixed polarization operator is still localized, although the corresponding decay length is longer, as discussed in Appendix~\ref{app:lengthscales}. 
In contrast to the rest space, the polarization operator of the low-energy bands displays slow decay, see Fig.~\ref{fig:SrVO3}(b). In addition to the slow decay of the absolute value, the sign of the individual matrix elements of the rest space polarization shows oscillations similar to Friedel oscillations and the RKKY interaction, which originate in the presence of a Fermi surface. 

This example illustrates that the electronic localization, and the resulting absence of non-local vertex corrections, is also applicable to the ``mixed'' diagrams that appear in the constrained RPA. 

\section{Conclusion and discussion}

We conclude that the RPA diagrams are dominant over vertex corrections if the gap is large so that RPA will accurately describe the dielectric function of these wide gap systems, at least at small frequency (see Appendix~\ref{app:frequency}). 
This relation between the size of the gap and the applicability of RPA diagrams is borne out by the comparison with experimental dielectric constants $\epsilon$ by Shishkin and Kresse~\cite{Shishkin07}: Good agreement between RPA and experiment is found in wide gap systems (MgO, C, LiF) and the largest disagreements occur for systems with a small gap (ZnO, GaAs, ZnS, CdS). A good agreement has also been found between the RPA and experimental molecular polarizabilities in aromatic molecules~\cite{Jorgensen20} (c.f., Appendix~\ref{app:benzene}), which also have a substantial gap.

Furthermore we have shown that the construction of the model Hamiltonian plays an important role in the applicability of the (c)RPA.
\emph{Ab initio} (c)RPA calculations based on the Kohn-Sham states and the Kohn-Sham gap $E_\text{KS}$, are ``blessed'' by a cancellation of the simplest local exciton binding vertex correction.
Using the true gap, $E_g$, in combination with the RPA leads, however, to poor results, as discussed in detail in Sec.~\ref{sec:local}.
Previous benchmarks of (c)RPA~\cite{Shinaoka15,Honerkamp18} used double-counting corrections to fix the positions of the bands in the single-particle spectrum (see, in particular, Appendix A of Ref.~\cite{Shinaoka15}), i.e., the true gap $E_g$. 

Demonstrating the applicability of (c)RPA to screening in real materials is especially timely since recent investigations~\cite{Kinza15,Shinaoka15,Honerkamp18,Han20} have (numerically) identified models where vertex corrections to cRPA are necessary.
These examples involved a Hubbard model for the rest space with intrinsically short-ranged interactions. Therefore, in those models the criterion that the propagation length is shorter than the interaction length cannot be fulfilled and non-local vertex corrections cannot be ignored compared to non-local Coulomb interactions. 
As we have shown, the assumption of local Hubbard interactions in the rest space is insufficient to understand screening in real materials.

We have shown that the cRPA is applicable to rest space bands far away from the Fermi level. 
A question that remains open is how to deal with rest space bands that are close to or even crossing the Fermi level. 
One important example is Nickel~\cite{Lichtenstein01,Braun06,Grechnev07,Miyake08,Miyake09,Vaugier11,SanchezBarriga12,Hausoel17}, where the $s$-band, which crosses the Fermi level, is often excluded from the low-energy model (see Fig.~1 of Ref.~\onlinecite{Miyake09}) and its screening is taken into account via cRPA. Our present work does not directly justify this approach. 
Further detailed investigation into the role of vertex corrections in this kind of system are warranted.

We should also mention a special example where RPA is known to be (surprisingly) close to numerically exact results, even though neither the dense electron gas limit nor the large band gap limit applies.
This is the electrostatic screening by $p_z$ electrons in undoped graphene~\cite{Astrakhantsev18}. The Dirac nature of the $p_z$ electrons could play a role here and length scale arguments similar to those presented in this work might provide a way to understand this computational result.

To summarize, we have proven that the (constrained) Random Phase Approximation is applicable to gapped systems and that vertex corrections vanish as the gap becomes large.
Due to the gap, the propagation length of electrons/holes is small compared to the length scale of the interaction responsible for the screening. 
The vertex corrections require quantum tunneling of particles and holes, with the gap acting as the tunneling barrier. 
The (c)RPA diagrams, on the other hand, involve only classical interactions between quantum fluctuations in the electronic density and can therefore screen over much larger length scales. 
The correspondingly larger phase space of the RPA processes means that non-local vertex corrections can be neglected, similar to Migdal's theorem for electron-phonon systems.
Finally, in gapped systems the most important local vertex corrections correspond to exciton binding. These effects are effectively included when using the Kohn-Sham states and Kohn-Sham gap in the (c)RPA calculation.

In addition to providing a justification for the use of cRPA to calculate effective interactions, our work shows how to establish a criterion for when to stop downfolding: The simple energetic picture of Fig.~\ref{fig:minimalmodel} breaks down when the orbital exchange interactions between valence and conduction states approach the magnitude of the gap (Fig.~\ref{fig:graphene:ed}).

\acknowledgments

The authors acknowledge useful discussions with Andy Millis and thank Merzuk Kaltak for sharing his cRPA routines with us. Financial support by the Zentrale Forschungsf\"orderung of the Universit\"at Bremen and the DFG via RTG 2247 QM$^3$ are acknowledged. 
The work of M.I.K. is supported by European Research Council via Synergy Grant 854843 - FASTCORR.

\bibliography{references}

\clearpage 
\appendix

\section{Length and Energy scales in a tight-binding model}

\label{app:lengthscales}

\begin{figure}
\includegraphics{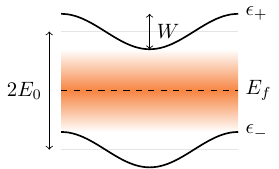}
\caption{Tight-binding model with a gap $\Delta=2E_0-W$.}
\label{fig:tightbinding}
\end{figure}

We now perform an explicit calculation of the length scales in a one-dimensional two-band tight-binding model (Fig.~\ref{fig:tightbinding}) with hopping $t>0$ and on-site energy $\pm E_0$, with $\abs{E_0} \gg t$. The bandwidth is $W=2t$ and the gap is $2E_0-W$.

\subsection{Green's function, real space}
In real space, the matrix Hamiltonian is tridiagonal:
\begin{align}
 \hat{H} &= 
 \begin{pmatrix}
  \ddots & \ddots & \ddots & 0 & \ldots \\
  0 & t & E_0 & t & 0 & \ldots \\
  \ldots & 0 & t & E_0 & t & 0 & \ldots \\
  &\ldots & 0 & t & E_0 & t & \ldots \\
  &&\ldots& 0& \ddots & \ddots & \ddots &
 \end{pmatrix} 
 \equiv E_0 \hat{\mathbb{I}} + t \hat{\mathbb{B}}
\end{align}
The Green's function is also a matrix in real space, 
\begin{align}
 \hat{G}(E) &= (E\, \hat{\mathbb{I}}-\hat{H})^{-1}  \\
 &= \frac{1}{E-E_0} \frac{\hat{\mathbb{I}}}{\hat{\mathbb{I}} - \frac{t}{E-E_0} \hat{\mathbb{B}}} \notag \\
 &\approx \frac{1}{E-E_0} \left(\hat{\mathbb{I}} + \frac{t}{E-E_0} \hat{\mathbb{B}} + \left(\frac{t}{E-E_0}\right)^2 \hat{\mathbb{B}}^2+\ldots  \right), \notag 
 \end{align}
 resulting in an exponential decay
\begin{align}
G(E,r-r') &\approx \frac{1}{E-E_0} \left(\frac{t}{E-E_0}\right)^{(r-r')/a} \label{eq:app:greenreal}
\end{align}
with decay length $\ell$,
\begin{align}
 \exp(-a/\ell) &= \abs{\frac{t}{E-E_0}} \notag \\
 \ell &= \frac{a}{\ln(\abs{\frac{E-E_0}{t}})}.
\end{align}

\subsection{Green's function, momentum space}

The same result can obtained starting from the dispersion and Green's function in momentum space,
\begin{align}
  E_\kv &= E_0  + 2t \cos(a k), \\
  G(E,\kv) &= \frac{1}{E-E_\kv}\notag
\end{align}
The electronic propagation length is determined by the real-space Green's function, so we perform a Fourier transform and expand in $t/(E-E_0)$:
\begin{align}
G(E,r) &= \frac{a}{2\pi} \int_{-\pi/a}^{\pi/a} dk \, G(E,k) \cos(k r) \\
  &= \frac{1}{E-E_0} \frac{a}{2\pi} \int_{-\pi/a}^{\pi/a} dk \frac{\cos(k r)}{1- \frac{2t}{E-E_0} \cos(ak)} , \notag \\
  G(E,r=0) &\approx \frac{1}{E-E_0} \notag \\
  G(E,r=a) &\approx \frac{1}{E-E_0} \frac{t}{E-E_0} \notag \\
  G(E,r) &\approx \frac{1}{E-E_0} \left(\frac{t}{E-E_0}\right)^{r/a}, \notag
\end{align}
as found before.
The exponential decay is typical for tunneling processes. 

The origin of this spatial decay is the offset $E-E_0$ between the argument of the Green's function and the band energy. If we instead consider $G(E=E_0,r)$, we find
\begin{align}
 G(E=E_0,r) &= \frac{a}{2\pi} \int_{-\pi/a}^{\pi/a} dk \frac{\cos(rk)}{2t \cos(ak)} \\
 &= \frac{1}{2t} \times
 \begin{cases}
 1 \text{ if $r=(2n+1)a$,} \\        
 -1 \text{ if $r=(2n+3)a$,} \\        
 0 \text{ if $r=(2n)a$ and $n\neq 0$} \\        
 \end{cases}.
\end{align}
Note that the integral is divergent at $r=0$, but we are interested in the behavior at large $r$, where this expression does not decay in magnitude at all. Thus, the localization of the Green's function $G(E,r)$ can only be considered once the energy argument $E$ is fixed, the localization occurs for particles that are ``off shell'', in quantum field theory terms. 

In particular, as a concept the localization of the Green's function is somewhat distinct from the localization of Wannier functions~\cite{Brouder07}, since this example shows that the latter is not a sufficient criterion for the former.

\subsection{Lindhard polarization}

We now consider a $d$-dimensional hypercubic lattice with lattice constant $a$ and two bands $E_+(\kv)$, $E_-(\kv)$. Both bands have the same bandwidth $W=4d\, t$, where $t$ is the hopping and the two bands have on-site energy $\pm E_0$ (this is the energy offset with respect to the Fermi level). Here, $2 E_0$ is the energy difference between the center of the two bands, the band gap is $\Delta=2E_0-W$. Explicitly, the dispersion relation is
\begin{align}
  E_{\pm} (\kv) &= \pm E_0  + 2t \sum_{i=1}^{d} \cos(a k_i) = \pm E_0+W f(a\kv). 
\end{align}
Here, the $+$ stands for the conduction band and the $-$ for the valence band and we have introduced the dimensionless function of order unity $f=\frac{1}{2d}\sum_{i=1}^{d} \cos(a k_i) $, with $\int d\kv f(a \kv)=0$. We again assume $t\ll E_0$, in other words, hopping smaller than the on-site energy.

We determine the exponential decay length of the polarization $\Pi$, using Lindhard's formula.
For notational convenience, we set $a=1$, and normalize $\int d\kv$ by the Brillouin Zone volume. Note that we do not include a spin factor 2 in the definition of $\Pi$.
The only allowed excitations (at zero temperature) are particle-hole pairs, i.e., the combination $\Pi_{+-}$. 
\begin{widetext}
\begin{align}
 \Pi_{+-}(\qv,\omega=0) &= \int d\kv \frac{1}{E_+(\kv+\qv)-E_-(\kv)} \\
 &= \int d\kv \, \frac{1}{2 E_0 + W\left[f(\kv+\qv)-f(\kv)\right] }  \\
 &= \frac{1}{2E_0} \int d\kv \, \frac{1}{1+\frac{W}{2 E_0} \left[f(\kv+\qv)-f(\kv)\right] } \\
 &\overset{W\ll \Delta}{\approx} \frac{1}{2 E_0} - \frac{W}{4 E_0^2} \int d\kv \left[f(\kv+\qv)-f(\kv)\right] + \frac{W^2}{8 E_0^3} \int d\kv \left[f(\kv+\qv)-f(\kv)\right]^2 +\ldots \label{eq:lengtscale1}\\
 &=\frac{1}{2 E_0} + 0 + \frac{W^2}{8 E_0^3} \frac{1}{2d^2} \sum_{i=1}^d  \sin\left( \frac{q_i}{2}\right)^2 \\
\Pi(\rv) &=  \int d\qv \, \Pi(\qv) \cos(\qv \cdot \rv) \\
\Pi(r=0) &\approx \frac{1}{2E_0} \\
\Pi(r=1) &\approx \frac{W^2}{32 d^2 E_0^3} 
\end{align}
\end{widetext}
The nearest-neighbor part of $\Pi$ is smaller than the local part by a factor $W^2/(16 d^2 E_0^2)=t^2/(E_0^2)$. This is the square of the decay of the Green's function, Eq.~\eqref{eq:app:greenreal}, since $\Pi$ is the product of two Green's functions. As a result, $\Pi$ decays exponentially with decay length $\lambda=\ell/2$. That $\Pi$ is largely local (momentum independent) was also observed in model studies by Honerkamp~\cite{Honerkamp12}.
Note that we have set $a=1$ and $\omega=0$ from the start in the calculation of $\Pi$. Expressing $\lambda$ in terms of $W$ and $d$ gives
\begin{align}
 \Pi(r) &\sim \exp(-2r/\ell) = \exp(-r/\lambda) \\
 \lambda &= \frac{a}{2 \ln\left( \frac{4 d E_0}{W} \right) }.
\end{align}
For SrVO$_3$, $a\approx 3.8$ \AA, $d=3$, $W\approx 5\,$eV and $E_0 = (\Delta+W)/2 \approx 4\,$eV, resulting in $\lambda=\ell/2=0.84$ \AA. Here, we used the bandwidth of the filled set of target bands, using $W=4$ eV results in $\lambda=0.8$ \AA, a rather similar result.

For the Green's function $G(E,r)$, localization occured if the energy $E$ was entirely ``off shell''. For the polarization, $\omega=0$ is the total energy carried by the particle-hole pair. Since $\omega$ is small compared to the gap, the particle and the hole cannot be simultaneously ``on shell'' and this guarantees the decay of the polarization.

In this calculation, the second term in  Eqn.~\eqref{eq:lengtscale1} vanished. This cancellation will occur for any dispersion relation $f$ with $\int d\kv f(\kv)=0$, as long as $E_0$ is sufficiently large to guarantee that the band does not cross the Fermi surface: in that case the integral is always carried out over the entire Brillouin Zone.

This evaluation of the polarization was done in the band basis, since Lindhard's formula can be used there. In calculations based on \emph{ab initio} bands, as presented in the main text, additional orbital overlap matrix elements of order unity play a role, but exponential decay is retained. The details of the dispersions of these bands determine the precise value of $\lambda$, the simple estimate given here based only on the energy scales provides the correct order of magnitude. For more details about transforming polarizations between band and orbitals basis sets, we refer the reader to Ref.~\cite{Kaltak}.

\subsection{RPA and vertex correction in tight-binding model}

The two second-order diagrams are shown in Fig.~\ref{fig:diagrams}. The second-order RPA diagram can be calculated directly in momentum space,
\begin{align}
 \chi^\text{RPA2}(\qv) \sim \frac{V_{\qv}}{4 E_0^2}, \label{eq:app:rpa2:tb}
\end{align}
which implies that the real space structure of $\chi^\text{RPA2}$ is determined by $V(r)$.

We now explicitly calculate the second-order vertex correction diagram for the tight-binding model, directly applying a lowest-order series expansion in $W/E_0$. We use that the interaction $V$ is instantaneous, i.e., independent of transferred energy.
\begin{widetext}
\begin{align}
 &\chi^{\text{VXC2}}(\rv_2 - \rv_1) \sim \sum_{\rv_3,\rv_4} \int dE_1 \int dE_2  G(\rv_4-\rv_1,E_1)G(\rv_3-\rv_1,E_1)G(\rv_2-\rv_4,E_2)G(\rv_3-\rv_2,E_2)V(\rv_4-\rv_3) \notag \\
 &\chi^{\text{VXC2}}\sim
 \sum_{\rv_3,\rv_4} \int dE_1 \int dE_2 
    \frac{t^{\abs{\rv_1-\rv_4}}}{(E_1+E_0)^{\abs{\rv_1-\rv_4}+1}}
    \frac{t^{\abs{\rv_3-\rv_1}}}{(E_1-E_0)^{\abs{\rv_3-\rv_1}+1}}
    \frac{t^{\abs{\rv_4-\rv_2}}}{(E_2+E_0)^{\abs{\rv_4-\rv_2}+1}}
    \frac{t^{\abs{\rv_2-\rv_3}}}{(E_2-E_0)^{\abs{\rv_2-\rv_3}+1}}
    V(\rv_4-\rv_3)
    \notag \\
&\chi^{\text{VXC2}}(\rv_2 - \rv_1)\sim \frac{V(\rv=0)}{4E_0^2} \delta_{\rv_2-\rv_1} + O\left(\frac{t^2}{E^4_0}\right) + \ldots \label{eq:app:vx2:tb} 
\end{align}
\end{widetext}

Comparing explicitly Eq.~\eqref{eq:app:rpa2:tb} and \eqref{eq:app:vx2:tb}, using $t/E_0 \approx 0$, the ratio of the second-order vertex correction and RPA diagrams is
\begin{align}
\left|\frac{\chi^{\text{VX2}}}{\chi^\text{RPA2}}\right|(\rv=0) &\overset{t/E_0\ll 1}{\rightarrow} 1, \\
\left|\frac{\chi^{\text{VX2}}}{\chi^\text{RPA2}}\right|(\rv\neq0) &\overset{t/E_0\ll 1}{\rightarrow} \frac{1}{V(\rv)}O\left(\frac{t^2}{E_0^2}\right) \overset{t/E_0\ll 1}{\rightarrow} 0
\end{align}
The second line shows that as long as $V(\rv)\neq 0$, non-local vertex corrections are small compared to non-local RPA diagrams in the large band gap limit.

It is important to state that electronic localization does not mean that the susceptibility is short-ranged. Indeed, Eq.~\eqref{eq:app:rpa2:tb} is long-ranged. Instead, only the irreducible part, $\Pi$, is short-ranged. This is somewhat similar to the Dynamical Mean-Field Theory limit~\cite{Georges96}, where the single-particle self-energy $\Sigma$ -- also an irreducible object -- becomes entirely local.

\subsection{Mixed polarization}

In the previous argument, both bands were far away from the Fermi level. The cRPA also considers the so-called rest-target polarization. In that case, one Green's function is close to the Fermi level and one is far away. We introduce an additonal band that crosses the Fermi level, with dispersion
\begin{align}
 h_0 = W f_0(\kv),
\end{align}
where $f_0$ is a dimensionless function of order unity with $\int d\kv f_0(\kv)=0$. 
and calculate the rest-target polarization, where we now have to take account of the range of the momentum integration, 
\begin{align}
 \Pi_{+0}(\qv) &= \int\limits_{\epsilon_0(\kv)<0} d\kv \frac{1}{E_+(\kv+\qv)-h_0(\kv)} \\
 &= \int\limits_{f_0(\kv)<0} d\kv \frac{1}{E_0+ W \left[ f(\kv+\qv) - f_0(\kv) \right] } \notag \\
 &=  \frac{1}{E_0} \int\limits_{f(\kv)<0} d\kv \frac{1}{1+ \frac{W}{E_0} \left[ f(\kv+\qv) - f_0(\kv) \right] } \notag \\
 &= \frac{1}{E_0} \left[1 - \frac{W}{E_0} \int\limits_{f_0(\kv)<0} d\kv \left(f(\kv+\qv)- f_0(\kv)\right) +\ldots \right].\notag
\end{align}
As in the rest-rest polarization, the leading order in the rest-target polarization is local (independent of $\qv$). However, unlike in Eq.~\eqref{eq:lengtscale1}, the first order term now does not vanish since the integral is no longer over the entire Brillouin Zone. As a result, the non-local part is suppressed by a facor $W/E_0$ only, instead of the $(W/E_0)^2$ found in the rest space.
Note that here, we also assumed that $W<E_0$, i.e., the bandwidth of the target space is also small compared to the gap.

\section{Dipolar nature of screening}
\label{app:dipolar}

Charge fluctuations are suppressed in gapped systems. As a result, the screening is not charge-like, as it would be in a metal. Instead, screening occurs via the excitation of particle-hole pairs, with the particle living in the conduction bands and the hole in the valence band. We perform a multipole expansion to show that the interaction between two particle-hole pairs is a dipole-dipole interaction.

Let $\Rv$ be a lattice vector with $\Rv=R\hat{\Rv}$, i.e., $R$ is the magnitude of $\Rv$ and $\hat{\Rv}$ is a unit vector. The RPA uses interactions between particles and holes, i.e., between bonding and antibonding orbitals, so we have $a\neq b$ and $c \neq d$ and calculate the matrix element
\begin{widetext}
\begin{align}
 V_{ab,cd}(\Rv) 
 &= \int d\rv d\rv' \frac{e^2}{\abs{\rv-\rv'-\Rv}} w^*_{a}(\rv) w_{b}(\rv) w^*_{c}(\rv') w_{d}(\rv') \\ 
 &\approx 
 \frac{e^2}{R} \overbrace{\langle a | b \rangle  \langle c | d \rangle}^{=0}+\frac{e^2 \hat{\Rv}}{R^2} \cdot (\langle a |\rv | b \rangle \overbrace{\langle c| d \rangle}^{=0} - \overbrace{\langle a | b \rangle}^{=0} \langle c|\rv | d \rangle ) -\frac{e^2}{R^3} \frac{3\hat{\Rv}_i \hat{\Rv}_j-\delta_{ij}}{2} \left[\langle a |\rv_i | b \rangle \langle c |\rv_j  | d \rangle+\langle a |\rv_j | b \rangle \langle c |\rv_i  | d \rangle\right] +O(R^{-4}). \notag 
 &\quad\quad \text{ (multipole expansion)} 
\end{align}
\end{widetext}
Here, for compactness, we have not written the vanishing terms containing $\langle a|b\rangle=0$ in the quadrupole term. If we now take $a=d$, $b=c$ and define $\cos \alpha = \hat{\Rv} \cdot \langle a|\rv|b\rangle$, we obtain
\begin{align}
    V_{ab,ba}(\Rv) = \frac{C_{dd}}{R^3} (1-3\cos^2 \alpha),
\end{align}
with $C_{dd}=e^2 \abs{\langle a|\rv |b\rangle}^2$. In other words, the interaction between two particle-hole pairs is a dipole-dipole interaction as long as the separation between the two pairs is large. This interaction decays slowly as a function of $\Rv$. In fact, the Fourier transform of the interaction between two dipoles with the same orientation $\dv$ is~\cite{Lahaye09}
\begin{align}
 V^{\text{dp-dp}}(\Qv) =& C_{dd} \left[\cos^2 \alpha_{\Qv} - \frac{1}{3}\right],
\end{align}
where $\alpha_{\Qv}$ is the angle between $Qv$ and $\langle a |\rv|b\rangle$.
Notably, the magnitude of the dipole-dipole interaction is independent of the absolute value of $\qv$, it only depends on the angle. The interaction stays finite close to $\qv=0$, but the angle $\alpha_{\qv,\dv}$ varies rapidly as one travels around $\qv=0$.

\section{Vertex correction versus RPA for electrons with short propagation length}

\label{app:explicitdiagrams}

The lowest order diagrams are given in Fig.~\ref{fig:diagrams}. 
Previously, we compared these two diagrams for a specific tight-binding model.
Here, we perform a comparison of the two diagrams in a continuum model, with the assumption that the electrons are strongly localized on a scale $\ell$.
We simplify the situation by taking $G(\rv)=0$ for $\abs{\rv}>\ell$ and $G(\rv)$ constant for $\abs{\rv} \leq \ell$, so that $\int d\rv G(\rv) \approx G(0) S_\ell$, with $S_\ell = \frac{4}{3} \pi \ell^3$ the volume of the sphere with radius $\ell$.

We start with the vertex correction.
At short distances, the dominant contribution is the particle-hole density-density interaction, with a contribution $1/\abs{\rv_3-\rv_4}$ and a corresponding integral over $\rv_3$ and $\rv_4$. 

To perform the spatial integrals, we first set $\rv_1=0$ by translation symmetry, so that three integrals remain. 
If we further set $\rv_4\approx \rv_1$, then any point $\rv_i$ within distance $\ell$ of $\rv_1$ is also within distance $\ell$ of $\rv_4$, so the second constraint can be dropped. For $\rv_4\neq\rv_1$, the value of the integral is always smaller, since the additional constraint reduces the integration volume, and the integrand has a definite sign.

This makes the spatial integrals tractable, it now comes down to the geometric problem of finding the average value of $1/\abs{\rv_2-\rv_3}$ for two random points in the unit sphere, which has $6/5$ as the answer.
Inserting the appropriate units, we find  
\begin{align}
 \chi^{\text{vx2}}(\rv_1,\rv_4) &\leq \chi^{\text{vx2}}(\rv_1,\rv_4=\rv_1) \notag \\
 &= \begin{cases}
\frac{6e^2}{5 \ell} G^4(0) S_\ell^2 \sim \ell^5 &\text{ for } \abs{\rv_1-\rv_4} \leq 2 \ell \\                                                                            
0 &\text{ otherwise}                                                                           \end{cases}\notag\\
 \chi(\qv)^{\text{vx2}} &\sim \frac{6e^2}{5 \ell} G^4(0)S_\ell^2 S_{2\ell} \sim \ell^8, \label{eq:chivx}
\end{align}
where the final equation is obtained by Fourier transformation, which involves the integration over the variable $\rv_4-\rv_1$, which has to lie in $S_{2\ell}$.

The simplest way to understand this result is that all three spatial integrals are restricted by $\ell$, giving $\ell^{3\times 3}$, and one internal interaction line provides $\ell^{-1}$ for a total of $\ell^8$. Any additional Coulomb vertex inserted provides two spatial integrals ($\ell^{2\times 3}$) and at best one Coulomb interaction ($\ell^{-1}$) for an addition overall factor $\ell^5$. 

The previous diagram is only relevant when $\rv_1$ and $\rv_4$ are close together, on the scale $\ell$. 
On the other hand, the RPA diagram is relevant for large distances: $\rv_2$ and $\rv_3$ can be far apart since the interaction is long-ranged.
We immediately go to momentum space since both the bubble and the dipole-dipole interaction (Appendix~\ref{app:dipolar}) are diagonal in $\qv$ and evaluate the RPA correction with the same number of vertices and Green's functions,
\begin{align}
 \chi^{\text{RPA2}}(\qv) &= \Pi(\qv)^2 \,\,V^\text{dp-dp}(\qv) \text{ with: }\\
 \Pi(\rv) &= \begin{cases}
-G^2(0) &\text{ for } \abs{\rv} \leq \ell \\                                                                            
0 &\text{ otherwise}                                                                           \end{cases}  \\
 \Pi(\qv) &\approx G^2(0) S_\ell \quad \text{   for } \ell q \ll 1 \\
 \chi^{\text{RPA2}}(\qv)&\approx C_{dd} G^4(0) S_\ell^2 \left[ \cos^2 \alpha_{\qv}-\frac{1}{3}\right]. \label{eq:chirpa2}
\end{align}
The contribution scales overall as $\chi^{\text{RPA2}} \sim \ell^6$.
In general, every higher order in RPA involves one interaction and one bubble and contributes
\begin{align}
 C_{dd} G^2(0) S_\ell \left[ \cos^2 \alpha_{\qv,\dv}-\frac{1}{3}\right].
\end{align}
The comparison of powers of $\ell$ shows that RPA bubbles are leading over the vertex correction in the limit $\ell\rightarrow 0$. Every bubble contributes $\ell^3$ and every vertex correction $\ell^5$.

\subsection{The proper limit $\ell\rightarrow 0$}

Naively, the formulas above imply no screening in the limit $\ell\rightarrow 0$. To obtain a non-trivial limit, one needs to use the normalization condition 
\begin{align}
    \int d\rv G(\rv)G(-\rv) \approx G(0)^2 S_\ell = \text{constant}.
\end{align}
In that case, all orders in the RPA are independent of $\ell$ and the vertex correction vanishes as $\ell^2$.

\section{Relation to Migdal's Theorem}
\label{app:migdal}

\begin{figure}
\includegraphics{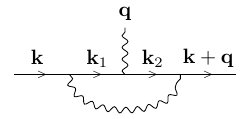}
\includegraphics{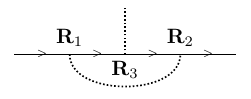}
\caption{Left: Migdal's theorem is based the fact that the energies $\epsilon(\kv_1)$ and $\epsilon(\kv_2)=\epsilon(\kv_1+\qv)$ have to be close to the Fermi level. This restricts the phase space for the momentum space integral $\int d\kv_1$.
Right: Our theory is based on the smallness of $\Rv_1-\Rv_3$ and $\Rv_2-\Rv_3$, which restricts the phase space of the real space integrals $\int d\Rv_i$.
}
\label{fig:migdal}
\end{figure}

A convenient statement of Migdal's theorem~\cite{Migdal58,Schrieffer18} is:
Due to the large difference in mass between the electron and the nuclei (i.e., the fact that $v_s/v_f$ is small), energy transfer between electrons and phonons is ineffecient, so all internal electronic propagators in diagrams need to have an energy close to the Fermi level. Here, the scale for being close in energy is set by the phonon properties, i.e., by the sound velocity $v_s$ (or the phonon frequency $\omega_d$). The theorem applies to metals, so it is natural to consider the implications of this in momentum space. A vertex correction is shown in Fig.~\ref{fig:migdal}. The momenta $\kv_1$ and $\kv_2$ both need to be close to the Fermi level in energy, this restricts the integrals over the internal momenta to a small region around the Fermi surface and suppresses the relative correction of the vertex correction diagram by a factor $\omega_D/E_F$.

Here, in a low-energy description of wide band gap systems, $\omega,W \ll \delta$, energy transfer via the Coulomb interaction is inefficient, so electronic excitations again need to involve a small energy transfer.
If an electron starts at location $\Rv_1$ then the end point of the propagator needs to be close to $\Rv_1$ as well. This restricts the integrals over real space in the vertex correction and suppresses their relative magnitude.

There are also differences.
First, in momentum space, there are many $\kv$ with the same energy (e.g., the Fermi surface), so it is possible to change $\kv$ without changing $\epsilon(\kv)$. 
Second, the Green's function is diagonal in momentum space (both end points have the same momentum $\kv$) and the proof of Migdal's theorem proceeds entirely in momentum space. The Green's function is not diagonal in real space and in analyzing RPA, it is useful to regularly switch between real and momentum space.
Third, a subtle aspect is the nature of the Coulomb interaction, since it has divergences as a function of $r$ or $q$. This favours diagrams where electrons are close together in real space, like the vertex correction diagram. For $d = 3$, this effect lowers the scaling by one power of $\ell$, resulting in a scaling $\ell^5$ instead of $\ell^6$ per vertex correction.

\section{Kohn-Sham states and finite temperature}
\label{app:temperature}

The way of fixing $E_a$ and $E_b$ and the determination of the Kohn-Sham gap can also be understood in the framework of a small, finite temperate. For convenience, we once again ignore the pair-hopping interaction $\Gamma_{abab}$, so that the states shown in Fig.~\ref{fig:minimalmodel} are the exact eigenstates.

There will be a finite thermal occupation of the $a$ orbital.
At low temperature, this density $\av{n_a}_\text{thermal}$ is determined entirely by the energy difference between the zero-temperature ground state \cOne\ and the lowest lying excited state with finite $n_a$, which is state \cFour. 
The auxiliary Kohn-Sham system 
$ H^\text{KS} = \sum_{\sigma} h_a n_{a,\sigma} + h_b n_{b,\sigma}$
has the correct density $\av{n_a}_\text{thermal}$ 
by construction and therefore needs to have the same energy difference 
$E^\text{KS}_\cFour - E^\text{KS}_\cOne = E_\cFour - E_\cOne$. 
This fixes $E_a$ and $E_b$ up to an overall additive constant.

\section{The role of frequency in screening}
\label{app:frequency}

Our attention has been focussed on static screening, i.e., $\omega=0$. In the cRPA, the screened interaction is, in general, retarded (a function of frequency)~\cite{Aryasetiawan04} and a relevant question is how well cRPA works at \emph{finite} frequency, $\omega>0$.
A physical interpretation of the frequency is easiest when we consider $\epsilon$ and $\chi$ as response functions, in that case $\omega$ is the frequency of the external field (photon energy). 

In a gapped system, $\omega=0$ implies that the particle-hole pair cannot be entirely on-shell. But once $\omega$ reaches the magnitude of the gap, the external field can excite on-shell particle-hole pairs away from the Fermi level and out of the classically forbidden zone and the present argumentation breaks down. 
The Kohn-Sham gap $E_\text{KS}$ is the relevant physical energy scale for the creation of particle-hole pairs and the RPA is expected to break down as $\omega \approx E_\text{KS}$.

When applied to downfolding, we should stress that it is not just the constrained Random Phase Approximation that breaks down. 
Instead the entire concept of downfolding is less applicable when the driving frequency $\omega$ is large. The high-energy properties of a system cannot be described by a low-energy downfolded model. For example, any downfolded model is unlikely to accurately describe the Electron Energy Loss Spectrum of the material at large $\omega$, since transition between higher energy states are explicitly relevant in that case. However, the goal of cRPA's $U(\omega)$ is to describe the feedback of (high-energy) excitations on the (low-energy) electronic spectrum and downfolding is suitable for that purpose.

\section{VASP calculations \label{app:vasp}}

All first principles calculations are performed within the Vienna Ab initio Simulation Package (VASP) \cite{kresse_ab_1993, kresse_efficiency_1996} using a PAW basis set \cite{kresse_ultrasoft_1999} and a GGA (PBE) exchange correlation functional \cite{perdew_generalized_1996}. All projections to localized Wannier orbitals are performed with the Wannier90 package \cite{pizzi_wannier90_2020} and all RPA and cRPA evaluations are done using a recent implementations by Kaltak within VASP \cite{Kaltak}. For the cRPA calculations we mostly use the projection-constraining scheme by Kaltak \cite{Kaltak} utilizing block-diagonalized projectors $\mathbf{U}(k)$. The later are defined as the rotation matrices transforming Kohn-Sham states $\Psi_n(k)$ to Wannier states $\phi_\alpha(k)$ according to
\begin{align}
    \phi_\alpha(k) = \sum_n U_{n\alpha}(k) \, \Psi_n(k)
\end{align}
and are the results from the Wannierization procedure \cite{Marzari12}. For graphene and SrVO$_3$ we separately construct three different projections $\mathbf{U}^{1,2,3}(k)$ for the valence, target, and conduction states and combine them afterwards.

    \subsection{Graphene}

    The graphene calculations are performed using an in-plane lattice constant of $a_0 = 2.468\,$\AA\, a super-cell height of $20\,$\AA\, applying an energy cut-off of about $515\,$eV, and using $18 \times 18 \times 1$ $k$/$q$ grids. The three distinct Wannier constructions for the bonding / anti-bonding and the $p_z$ states were performed individually using different initial projections and wannierization windows. The bonding orbitals ($b$) were constructed from three initial $s$-shaped orbitals localized at the C-C bond center and using a ``frozen'' (inner) wanierization window including all Kohn-Sham states between $-20\,$eV and $-2.4\,$eV below the Fermi level. The $p_z$ orbitals were constructed using $p_z$ initial projections centered on each C atom and using a ``frozen'' window from $-2.6\,$eV to $1.6\,$eV. The anti-bonding orbitals ($a$) were constructed from initial in-plane $p_x$ and $p_y$ orbitals positioned at the C-C bond centers. Due to the entanglement of these states with other ``rest'' and the $p_z$ states (see Fig.~\ref{fig:graphene}) we cannot use a ``frozen'' window and just restricted the wannierization to an ``outer'' window spanning from  $2.6\,$eV to $25.5\,$eV above the Fermi level. All Wannier constructions were maximally localized. For all cRPA calculations we used in total $192$ bands with energies up to about $82\,$eV above the Fermi level.
    
    From these three individual Wannier constructions we gain three distinct projections (rotation matrices) $\mathbf{U}^{b,p_z,a}(k)$ which we use to construct three individual Wannier Hamiltonians $\mathbf{H}^{b,p_z,a}$. From the later we construct all needed non-interacting polarizations for the various sub-spaces and with the combined $\mathbf{U}(k)$ (from all three $\mathbf{U}^{b,p_z,a}$) we can calculte $U^{T1}$ and $U^{T2}$ from first principles within VASP. For the minimal localized bonding/anti-bonding model from Sec.~\ref{sec:BABmodel} we use local on-site energies $h_b = \mathbf{H}_{00}^b(R=0)$ and $h_a = \mathbf{H}_{00}^a(R=0)$ of a single bonding/anti-bonding orbital pair and define $U_b=U_{bbbb}^{T1}(R=0)$,  $U_a=U_{aaaa}^{T1}(R=0)$, $V_{ab}=U_{aabb}^{T1}(R=0)$, and $J_{ab}=U_{abab}^{T1}(R=0)$ with $a$ and $b$ corresponding the Wannier orbitals 3 and 6 depicted in Fig.~\ref{fig:graphene:wannier}.
    
    \subsection{SrVO$_3$}
    
    For SrVO$_3$ we use the structure from Ref.~\onlinecite{SrVO3_structure_2014} with a cubic cell and a lattice constant of $a_0 \approx 3.9\,$\AA. The DFT calculations are performed using a $5 \times 5 \times 5$ $k$ grid and an energy cut-off of $500\,$eV. As in the case of graphene, we construct the polarization from a block-diagonalized tight-binding Hamiltonian which we gain from three individual Wannier constructions for the predominantly O $p$ valence band ($-2$ to $-7\,$eV below the Fermi level), the metallic V $t_{2g}$ band ($-1$ to $+1\,$eV around the Fermi level), and the V $e_g$ conduction band ($+1$ to $+5\,$eV above the Fermi level). While the O $p$ and V $t_{2g}$ bands are not entangled with other Kohn-Sham states the V $e_g$ bands overlap with some higher ``rest'' bands. Thus, in the latter case we cannot use a full ``frozen'' Wannier window and restrict it from $+1.4\,$ to $+3.6\,$eV above the Fermi level. All Wannier functions are maximally localized and we use $64$ bands with energies up to about $+40\,$eV above the Fermi level to evaluate the back ground screening.
    
    \subsection{Benzene}
    
    We set the C-C distance to $1.395\,$\AA\ and relax the C-H distance yielding a separation of $1.086\,$\AA. For all calculation we use an energy cut-off of about $517\,$eV and utilize just a single $k$ point. The super-cell box size is varied as indicated in section \ref{app:benzene}. We define the correlated sub-space as those six Kohn-Sham states with the strongest C $p_z$ character around the Fermi level. Correspondingly we construct the localized Wannier orbitals using $p_z$ orbitals on the C sites as initial projections, which we maximally localized. The ``frozen'' window is including all states between $-3.5$ and $+5.0\,$eV around the Fermi level. Here, we use for the cRPA calculations the weighted constraining scheme by Friedrich et al. \cite{Sasioglu11}.

\section{Benzene as a benchmark}
\label{app:benzene}

\begin{figure}
\includegraphics{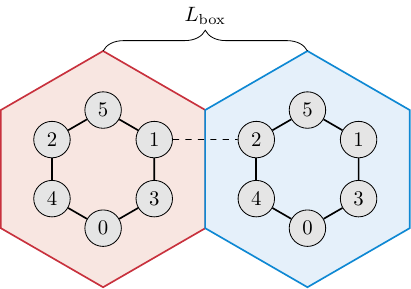}
\caption{Finite simulation box size effects in the benzene molecule (not to scale). The red box is the simulation box and the blue box is one of its periodic images. Since VASP calculations use periodic boundary conditions, matrix elements between, e.g., atoms 1 and 2 include unphysical contributions between periodic images (dashed line). This fictitious effect vanishes in the limit $L_\text{box}\rightarrow \infty$.}
\label{fig:benzene:box}
\end{figure}

In the two-tiered downfolding, for computational reasons we had to make the approximation that the Coulomb interaction depends only on a single momentum, i.e., $V(\Qv,\Kv,\Kv')\approx V(\Qv)$. In this way, some matrix elements with orbitals in different unit cells are neglected. In solids, the quality of this approximation depends on the particular Wannier construction and how much spillage it has. 

When simulating a single molecule, on the other hand, there should formally not be any spillage since there is only a single ``unit cell''. In practical VASP calculations, however, a finite simulation box of size $L_\text{box}$ with periodic boundary conditions is used, as in Fig.~\ref{fig:benzene:box}. In our two-tiered RPA, this lead to deviations between the direct VASP RPA calculation and the two-tiered approach. This error is controlled by the parameter $L_\text{box}$. Below, we study this effect quantitatively.

As in the main text, we compare a two-tiered approach to a direct calculation of screened Coulomb matrix elements between $p_z$ orbitals. In this case, tier 1 correspond to a cRPA calculation with all six $p_z$ orbitals frozen, i.e., in the target space. Tier 2 represents a full RPA calculation without any frozen orbitals (no target space). We compare the matrix elements between the $p_z$ orbitals, which are numbered as in Fig.~\ref{fig:benzene:box}. In Table~\ref{table:benzene:density}, we show the density-density elements for several sets of computational parameters. As the box becomes larger, the deviation between the direct and the two-tiered calculations becomes smaller, as anticipated. Similarly, increasing the number of Kohn-Sham states $N_B$ reduces the deviation. In Table~\ref{table:benzene:all}, we show all Coulomb matrix elements for $L_\text{box}=10.9$ \AA and $N_B=128$, showing that the density-density elements are largest in magnitude and also feature the largest deviation. In fact, the largest deviation occurs for the interaction between opposite sides of the molecule (orbitals 0 and 5), since the atom in the next periodic image is relatively close in that case. The Tables show that the deviations can be decreased by improving the computational parameters and the deviations are substantially smaller than the relevant matrix elements. 

In Table~\ref{table:benzene:all}, we should also point out that several RPA matrix elements are larger in magnitude than their cRPA counterparts, i.e., anti-screening occurs. 

\clearpage
\begin{table}
 \begin{tabular}{lr|ccc|ccc|ccc|ccc}
 \hline \hline
 &&& $U_0$ &&& $U_1$ &&& $U_2$ &&& $U_3$ \\
 $L_\text{box}$ & $N_B$&VASP&2tier&diff&VASP&2tier&diff&VASP&2tier&diff&VASP&2tier&diff \\
  \hline
  $7.25$ & $64$  &  7.7277 &  8.0523 & -0.3246&  5.6403 &  6.0021 & -0.3618&  4.9622 &  5.3835 & -0.4214&  4.9777 &  5.4296 & -0.4519 \\
  \hline
  $10.9$ & $64$ &  8.0312 &  8.1272 & -0.0960&  5.8787 &  6.0000 & -0.1213&  5.1709 &  5.3288 & -0.1579&  5.2207 &  5.3977 & -0.1770 \\
  $10.9$ & $128$ &  7.7102 &  7.7850 & -0.0747&  5.7685 &  5.8681 & -0.0996&  5.1064 &  5.2428 & -0.1364&  5.1119 &  5.2675 & -0.1556 \\
  \hline
  $14.5$ & $64$  &  8.6224 &  8.6552 & -0.0328&  6.1178 &  6.1713 & -0.0536&  5.3252 &  5.4026 & -0.0774&  5.4208 &  5.5109 & -0.0900\\
  $14.5$ & $128$ &  8.1995 &  8.2165 & -0.0170&  5.9253 &  5.9630 & -0.0378&  5.1934 &  5.2562 & -0.0629&  5.2370 &  5.3132 & -0.0762 \\
  $14.5$ & $192$ &  8.0334 &  8.0478 & -0.0144&  5.8817 &  5.9163 & -0.0347&  5.1767 &  5.2363 & -0.0596&  5.1972 &  5.2701 & -0.0729 \\
  \hline \hline
 \end{tabular}
 \caption{On-site, nearest-neighbor, next-nearest-neighbor and next-next-nearest-neighbor matrix elements in benzene in eV. The three numbers denote the direct VASP RPA calculation, the two-tiered approach and their difference, respectively. $L_\text{box}$ is the in-plane length of the computational cell in \AA, the out-of-plane size of the box is held constant at $9.8$ \AA. $N_B$ the number of bands in the calculation.}
 \label{table:benzene:density}
\end{table}

\clearpage

\begin{table}
\small
\centering
\begin{tabular}{cccc|ccc|c}
\hline \hline
i & j & k & l & VASP cRPA & VASP RPA & two-tiered RPA & diff\\
\hline
0 & 0 & 0 & 0 &  9.3826 &  7.7102 &  7.7850 & \cellcolor{ErikRood}{-0.0747} \\
0 & 0 & 0 & 1 & -0.3177 & -0.2611 & -0.2597 & {-0.0015} \\
0 & 0 & 0 & 3 &  0.5317 &  0.3987 &  0.3954 & { 0.0033} \\
0 & 0 & 0 & 5 &  0.2408 &  0.2221 &  0.2219 & { 0.0001} \\
0 & 0 & 1 & 1 &  4.7005 &  5.1064 &  5.2428 & \cellcolor{ErikRood}{-0.1364} \\
0 & 0 & 1 & 2 & -0.0654 & -0.1271 & -0.1296 & { 0.0025} \\
0 & 0 & 1 & 3 &  0.1483 &  0.1865 &  0.1866 & {-0.0001} \\
0 & 0 & 1 & 4 &  0.1477 &  0.1570 &  0.1571 & {-0.0001} \\
0 & 0 & 1 & 5 &  0.0874 &  0.1821 &  0.1853 & {-0.0032} \\
0 & 0 & 3 & 3 &  6.1192 &  5.7685 &  5.8681 & \cellcolor{ErikRood}{-0.0996} \\
0 & 0 & 3 & 4 & -0.2676 & -0.2210 & -0.2187 & {-0.0024} \\
0 & 0 & 3 & 5 & -0.1009 & -0.1499 & -0.1512 & { 0.0014} \\
0 & 0 & 5 & 5 &  4.2522 &  5.1119 &  5.2675 & \cellcolor{ErikRood}{-0.1556} \\
0 & 1 & 0 & 1 &  0.0385 &  0.0319 &  0.0318 & { 0.0002} \\
0 & 1 & 0 & 2 &  0.0184 &  0.0191 &  0.0192 & {-0.0001} \\
0 & 1 & 0 & 3 & -0.0421 & -0.0351 & -0.0349 & {-0.0002} \\
0 & 1 & 0 & 4 & -0.0187 & -0.0178 & -0.0178 & {-0.0000} \\
0 & 1 & 0 & 5 & -0.0198 & -0.0195 & -0.0195 & {-0.0000} \\
0 & 1 & 2 & 3 & -0.0155 & -0.0160 & -0.0160 & { 0.0000} \\
0 & 1 & 2 & 4 & -0.0103 & -0.0182 & -0.0184 & { 0.0002} \\
0 & 1 & 3 & 4 &  0.0207 &  0.0203 &  0.0202 & { 0.0001} \\
0 & 1 & 4 & 5 &  0.0101 &  0.0162 &  0.0164 & {-0.0002} \\
0 & 3 & 0 & 3 &  0.1758 &  0.1512 &  0.1509 & { 0.0003} \\
0 & 3 & 0 & 4 &  0.0179 &  0.0160 &  0.0159 & { 0.0002} \\
0 & 3 & 0 & 5 &  0.0265 &  0.0253 &  0.0253 & { 0.0000} \\
0 & 3 & 1 & 4 &  0.0197 &  0.0221 &  0.0221 & {-0.0000} \\
0 & 3 & 1 & 5 &  0.0213 &  0.0288 &  0.0289 & {-0.0002} \\
0 & 3 & 2 & 5 &  0.0155 &  0.0289 &  0.0292 & {-0.0003} \\
0 & 5 & 0 & 5 &  0.0221 &  0.0209 &  0.0209 & { 0.0000} \\
0 & 5 & 1 & 4 &  0.0152 &  0.0158 &  0.0158 & {-0.0000} \\
\hline \hline
\end{tabular}
\caption{Coulomb matrix elements in benzene. All numbers in eV. Matrix elements with a deviation $>0.01\,$eV between the two-tiered RPA and direct VASP RPA are highlighted in red. Only distinct matrix elements are shown. All other orbital combinations can be recovered by symmetry. These data correspond to $L_\text{box}=10.9$ \AA and $N_B=128$, see also Table~\ref{table:benzene:density}.}
\label{table:benzene:all}
\end{table}

%

\end{document}